# Pressure Induced Collapse of Magnetic Order in Jarosite

Ryan A. Klein,[1] James P. S. Walsh,[1] Samantha M. Clarke,[2] Zhenxian Liu,[3] E. Ercan Alp,[4] Wenli Bi,[5] Yue Meng,[6] Alison B. Altman,[1] Paul Chow,[6] Yuming Xiao,[6] M. R. Norman,[7,†] James M. Rondinelli,[8,†] Steven D. Jacobsen,[9,†] Danilo Puggioni,[8,†] and Danna E. Freedman[1,†]

[1] *Department of Chemistry, Northwestern University, Evanston, Illinois 60208, United States*
[2] *Lawrence Livermore National Laboratory, 7000 East Ave, Livermore, California 94550, United States*
[3] *Department of Physics, University of Illinois at Chicago, Chicago, Illinois 60607, United States*
[4] *Advanced Photon Source, Argonne National Laboratory, 9700 South Cass Avenue, Lemont, Illinois 60439, United States*
[5] *Department of Physics, University of Alabama at Birmingham, Birmingham, Alabama, 35294, United States*
[6] *HPCAT, X-Ray Science Division, Argonne National Laboratory, Lemont, Illinois 60439, United States*
[7] *Materials Science Division, Argonne National Laboratory, Argonne, Illinois 60439, United States*
[8] *Department of Materials Science and Engineering, Northwestern University, Evanston, Illinois 60208, United States*
[9] *Department of Earth and Planetary Sciences, Northwestern University, Evanston, Illinois 60208, United States*

We report a pressure-induced phase transition in the frustrated kagomé material jarosite at ~45 GPa, which leads to the disappearance of magnetic order. Using a suite of experimental techniques, we characterize the structural, electronic, and magnetic changes in jarosite through this phase transition. Synchrotron powder X-ray diffraction and Fourier transform infrared spectroscopy experiments, analyzed in aggregate with the results from density functional theory calculations, indicate that the material changes from a $R\bar{3}m$ structure to a structure with a $R\bar{3}c$ space group. The resulting phase features a rare twisted kagomé lattice in which the integrity of the equilateral $Fe^{3+}$ triangles persists. Based on symmetry arguments we hypothesize that the resulting structural changes alter the magnetic interactions to favor a possible quantum paramagnetic phase at high pressure.

Spins arrayed on lattices exhibiting magnetic frustration can engender exotic magnetic phases [1–4]. Materials that host these lattices have been intensely researched for several decades, including the antiferromagnetic mineral jarosite, $KFe_3(OH)_6(SO_4)_2$ [5–15]. Jarosite's kagomé lattice – in which high-spin $Fe^{3+}$ ions ($S=\frac{5}{2}$) form a corner-shared triangular lattice – frustrates magnetic ordering. The Néel temperature, $T_N$, of jarosite at ambient pressure is 65 K – much lower than may be expected given the Curie-Weiss temperature of −828 K, leading to a frustration parameter of f = 12.7, where f ≥ 10 is considered frustrated [9,16,17].

The ordered magnetic structure of jarosite arises from two primary interactions [10–14]. The first is a nearest neighbor antiferromagnetic superexchange interaction along the Fe–O–Fe pathway within the kagomé lattice. The second is an antisymmetric Dzyaloshinskii-Moriya (DM) interaction [18,19]. The DM vectors, **D**, are confined within the mirror plane bisecting the $Fe^{3+}$ ions [12,18–21]. These interactions select the q=0 spin structure which features a magnetic umbrella motif with a uniform positive vector chirality and a small canting of the spins out of the plane [12,13,14]. This canting alternates from plane to plane, but an applied magnetic field can align them leading to a net ferromagnetic moment [12].

Extensive theoretical work hints at a rich phase diagram for frustrated kagomé antiferromagnets in the presence of a DM interaction [12,22–30]. These studies predict quantum critical points between the q=0 state and other magnetic phases that are potentially accessible in jarosite by manipulating the exchange interactions. To that end, high applied pressures offer a vector to tune the phase stability [31–35], structure, and magnetism in frustrated materials analogous to variable magnetic fields and chemical composition. High applied pressures shorten interatomic distances, which impact the potential energy landscapes and magnetic exchange pathways for the realization of exotic phases of matter [36–41]. Moreover, applied pressure also offers a route to access magnetically frustrated variants of parent lattices that may be difficult to access synthetically at ambient conditions, which may also host exotic magnetic phases. Despite the growing body of experimental work in this area, complete *PT* phase diagrams for kagomé lattice materials– and other frustrated lattices featuring competing exchange and DM interactions – remain absent.

Here, we report the high-pressure quenching of magnetic order in jarosite leading to a possible quantum paramagnetic phase. We study the crystal, electronic, and magnetic structure of jarosite at pressures up to 121 GPa using a suite of *in situ* diamond anvil cell (DAC) techniques [42]. We identify a symmorphic-to-nonsymmorphic transition from $R\bar{3}m$ to $R\bar{3}c$. The calculated $R\bar{3}c$ structure exhibits equilateral triangles of $Fe^{3+}$ ions that twist, but do not distort, to yield a rare twisted kagomé lattice with a unit cell that doubles along the *c*-axis. This conclusion is supported by a Rietveld refinement analysis of a high-pressure PXRD pattern. The lowering in symmetry is supported by Fourier-transform infrared spectroscopy (FTIR) measurements. Variable-temperature and -pressure synchrotron Mössbauer spectroscopy (SMS) measurements reveal a collapse of detectable magnetic order at pressures above 43.7(4) GPa down to 20 K, which we discuss in the context of the pressure-tuned exchange interactions on the twisted kagomé lattice to conjecture the existence of a quantum paramagnetic phase.

To search for potential pressure-driven phase transitions in jarosite, we collected powder X-ray diffraction (PXRD) patterns up to 78.6 GPa at ambient temperature. We extracted the unit cell lattice parameters at each pressure by applying a Pawley fit to the diffraction patterns (Fig. 1, top) [43–45]. Pawley fits using the ambient-pressure $R\bar{3}m$ structure model the data well across the entire measured pressure range. Up to 19.5(1) GPa, an initial third-order Birch Murnaghan equation-of-state (BM3) curve accurately models the data (Table S3) [46,47]. There is an isosymmetric transition at $P^{*,1}$=19.5(1) GPa, discernable from the inflection point in the $c/a$ ratio at this pressure (Fig. 1, bottom). Between 19.5 and 39.4 GPa, a second BM3 equation of state curve accurately models the data. We find a transition at $P^{*,2}$=43.7(4) GPa where an unusual cusp in $c/a$ occurs, signaling a complicated structural transition at this pressure. From 49.9 GPa to 78.6 GPa, the data are well modelled with a third BM3 equation of state curve.

To better understand this complex behavior under compression, we conducted DFT calculations to search for possible pressure-induced lattice instabilities. We constrained the volume of the unit cell to the observed volume at a given pressure and calculated the phonon modes every ten GPa up to 80 GPa. We find a soft phonon at the T (0,0,3/2) point of the Brillouin zone between 30 and 40 GPa (Fig. S31). Upon condensing the instability followed by atomic relaxation, we find a new high-pressure structure ($R\bar{3}c$ space group) which emerges from the known ambient pressure $R\bar{3}m$ structure. The two space groups have a group-subgroup relationship such that the transition is driven by a $T_2^+$ mode, which doubles the length of the $c$-axis due to the glide operation. The new, high-pressure $R\bar{3}c$ structure is characterized by a twisted kagomé lattice in which uniform Fe–Fe contacts form equilateral triangles that twist slightly relative to the parent kagomé lattice. At the same time, both apical oxygen ions shift in the same direction, enhancing the distortion of the FeO$_6$ octahedra. We distinguish this lattice from the highly distorted triangular-kagomé lattice [48], in which next near neighbor magnetic ions are connected by direct superexchange pathways. We analyzed the PXRD patterns over the entire measured pressure range considering both the $R\bar{3}m$ and $R\bar{3}c$ structures. We found that the Pawley fits of the patterns above the phase transition using the calculated $R\bar{3}c$ structure give comparable fit statistics to the Pawley fits which used the $R\bar{3}m$ structure. Additionally, the normalized lattice parameters qualitatively agree well with those obtained from the previous Pawley fits (Figs. S7–S10).

To investigate the possible lowering in symmetry at $P^{*,2}$, we conducted ambient temperature FTIR measurements in the mid-IR (500–6000 cm$^{-1}$) up to 54.4 GPa (Figs. S12–S21) [49–52]. At 45.5(4) GPa, there is a first order discontinuity in the IR modes. We observe an increase in the number of observed modes in the measured region in agreement with the calculated lowering in symmetry at the $R\bar{3}m$ to $R\bar{3}c$ transition. The calculated number of IR active modes exceeds the observed number of modes at all measured pressures and there is good qualitative agreement for the pressure-dependent dispersion between the calculated and the observed modes (Figs. S20, S21). These data support the DFT predicted symmorphic-to-nonsymmorphic transition.

To test the validity of the calculated structures at P>$P^{*,2}$, we performed a Rietveld refinement analysis of a selected representative PXRD pattern collected at 62.1(6) GPa. The details of the Rietveld refinement and structure solution are given in the supplemental document (pages S9−S11), and the results are summarized in Table S7. The Rietveld refinement using the $R\bar{3}c$ structure fits the experimental PXRD pattern well (fit statistics are listed in Table S7). In addition, the $R\bar{3}c$ structure is internally consistent with the FTIR measurements, while the other structures suggested by the DFT calculations are not. Therefore, based on the Rietveld refinement fits – analyzed in conjunction with the FTIR results and DFT calculations – we conclude that the high-pressure phase of jarosite is accurately described by the $R\bar{3}c$ structure and possesses a twisted kagomé lattice.

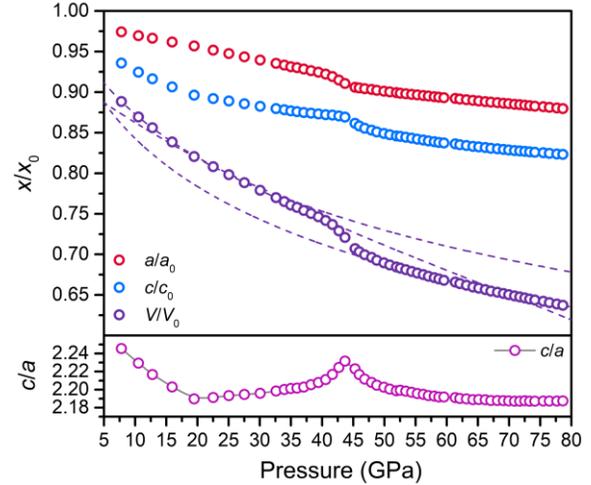

**FIG. 1** | The normalized lattice parameters for jarosite at ambient temperature are plotted as a function of pressure (top). These values were obtained from Pawley fits of the PXRD data using the $R\bar{3}m$ space group for the jarosite phase. The plot of $c/a$ vs pressure highlights the phase transitions at $P^{*,1}$ and $P^{*,2}$. The spline interpolation in the plot of $c/a$ vs pressure, which lies behind the data points over the entire measure pressure range, is a guide for the eye. Unless shown, error bars are commensurate with the symbol size. The dashed lines in the figure (top) are the normalized equation of state curves for the volume data (see Table S3 for parameters and Figs. S3–S6).

To probe the local electronic structure of the Fe ions in the high-pressure phase (P>$P^{*,2}$), we conducted non-resonant X-ray emission spectroscopy (XES) measurements as well as ambient-temperature, variable-pressure SMS measurements on pure single crystals of jarosite. Fig. 2 (left) summarizes the XES experiments. Up to 75 GPa, the K$\beta_{1,3}$ line red shifts slightly and decreases in intensity. The K$\beta'$ feature decreases in intensity slightly across this same pressure interval but does not shift in energy. We quantified the extent of the spectral changes as a function of pressure using the integral of the absolute value of the difference of the curves method (IAD) [53] (Fig. 2, left, inset). The IAD values follow a linear trend with pressure ($R^2 = 0.9556$). The absence of discontinuities in the plot of the IAD values vs. pressure confirms that there are no phase transitions in the local electronic structure of the Fe$^{3+}$ ions. Based on these findings, we conclude that the spin state of the Fe$^{3+}$ ions ($S$=5⁄2) remains constant up to 75 GPa [54–56]. We attribute the observed spectral changes to an increase in the covalent nature of the Fe–O bonds in the distorted FeO$_6$ octahedra in the kagomé planes [54,55,57]. Additionally, we reference our data in Fig. 2 with previously published data for compounds with established valence and spin states, including hematite, Fe$_2$O$_3$ (high-spin Fe$^{3+}$), measured at ambient conditions and phase D, MgSi$_{1.5}$Fe$_{0.15}$Al$_{0.32}$H$_{2.6}$O$_6$ (low-spin Fe$^{3+}$), measured at 93 GPa and ambient temperature [54], to show that no spin-state transition occurs up to 75 GPa. We also use reported data for Mg$_{0.75}$Fe$_{0.25}$O measured at 90 GPa as a low-spin Fe$^{2+}$ reference [58].

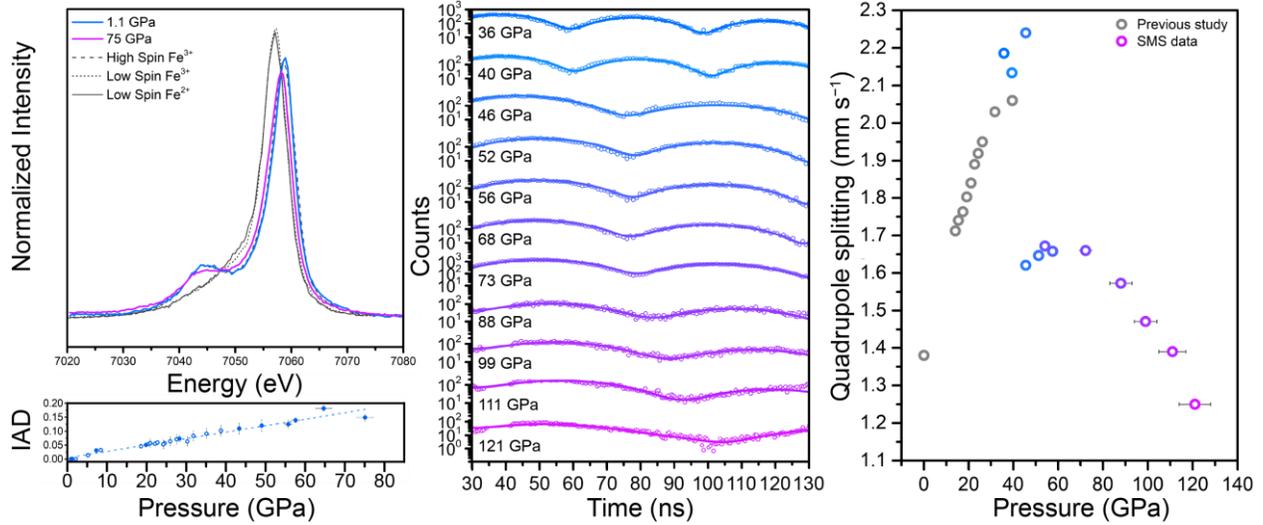

**FIG. 2** | The normalized XES data (left) illustrate that the electronic structure changes linearly with pressure, as shown by the plot of the IAD value vs. pressure in the inset of the XES plot. The data from jarosite measured at 1.1 GPa overlays the high-spin $Fe^{3+}$ reference data. In the plot of the IAD values, closed symbols originate from spectra reported herein, whereas open symbols originate from previously reported spectra [42]. The SMS data (center) show a change in spectral shape between 39.5 and 45.6 GPa. Spurious signals in the SMS data above 80 GPa are masked between 36 and 41 ns. Open circles denote the data points while the solid lines show the fits to these data. Fits of the data reveal a first order discontinuity in $\Delta E_Q$ (right), which decreases from 2.24(2) to 1.62(2) mm s$^{-1}$ through the phase transition at 43.7(4) GPa. Unless shown, error bars are commensurate with the symbol size. Open grey circles are previously reported data for the $\Delta E_Q$ plot [42]. See Fig. S23 for an enlarged plot of the IAD values as a function of pressure.

We further probed the local electronic and structural environment around the Fe$^{3+}$ ions using SMS experiments [59–64] at ambient temperature up to 121 GPa (Fig. 2, center). This technique is analogous to ambient pressure, offline Mössbauer experiments, and as such, yields the quadrupole splitting and magnetic hyperfine terms at elevated pressures. We fit these data to extract the quadrupole splitting values ($\Delta E_Q$; Fig. 2, right) [65]. $\Delta E_Q$ increases linearly with pressure from ~1.4 mm s$^{-1}$ at ambient pressure to 2.13(2) mm s$^{-1}$ at 39.5 GPa [42]. The spectrum at 45.6 GPa was fit with two phases, one with $\Delta E_Q$ = 2.24(2) mm s$^{-1}$ and one with $\Delta E_Q$ = 1.62(2) mm s$^{-1}$. At this pressure, there is a first order discontinuity in $\Delta E_Q$. Above the pressure-induced discontinuity, $\Delta E_Q$ remains relatively constant between 1.6 and 1.7 mm s$^{-1}$ up to 72.3 GPa and then decreases monotonically with additional pressure [66,67].

The magnitude of $\Delta E_Q$ stems from the asymmetry of the electric field gradient around the Mössbauer active ion [68,69]. In general, $\Delta E_Q$ is influenced by two factors: the symmetry of the electronically populated orbitals and the lattice contribution. In jarosite ($d^5$, high spin, nominal $^6S$ ground state), the non-zero value of $\Delta E_Q$ comes from the FeO$_6$ tetragonal elongation and from the differences in the σ- and π-donor ability between the equatorial μ$_2$–OH ligands compared to the axial sulfate ligands. One explanation for this behavior would be a spin crossover transition, however, within the resolution of the XES experiments, the IAD analysis shows no discontinuities in the electronic structure of the Fe$^{3+}$ ions up to 75 GPa. The combination of these data suggests the discontinuity in the $\Delta E_Q$ vs. pressure relationship arises from the sudden change in the lattice contribution to the asymmetry of the electric field gradient. The first order discontinuity in the plot of $\Delta E_Q$ vs. pressure therefore supports the calculated structural transition from the $R\bar{3}m$ to the $R\bar{3}c$ phase.

We probed the magnetic ordering temperature in jarosite by fitting the SMS data and extracting the magnetic hyperfine term, $B_{HF}$. $B_{HF}$ is the magnitude of Zeeman splitting of the $m_I$ sublevels probed in Mössbauer spectroscopy [68,69]. Here, a non-zero $B_{HF}$ arises from magnetic ordering. A non-zero $B_{HF}$ term creates readily apparent additional quantum beats in the SMS spectra that occur with greater frequency, and the lack of the onset of magnetic ordering as a function of pressure in the data presented in Fig. 2 is clear from the lack of additional quantum beats [42, 59−64, 68, 69]. Further details concerning the fitting of the data to extract the relevant Mössbauer terms can be found in the supplemental document. From these data, we extend the temperature-pressure magnetic phase diagram for jarosite to above 100 GPa in pressure (Fig. 3). $T_N$ for jarosite increases linearly with pressure up to ~40 GPa. This trend can be explained by considering the pressure-induced change in the equatorial Fe–O bond distances, d(Fe–O)$_{eq}$. This bond distance decreases with increasing pressure [42]. This affects both the exchange interactions $J$ and $\mathbf{D}$ as the Fe-3d to O-2p hopping integral $t_{pd}$ typically scales as $d(Fe-O)_{eq}^{-4}$. These interactions combine to yield a linear increase in $T_N$ up to ~40 GPa. Then, the variable-temperature, isobaric data set measured at ~47 GPa – just above $P^{*,2}$ – exhibits a $B_{HF}$ value of zero for all measured temperatures down to the lowest measured temperature of 29.3 K. The disappearance of the $B_{HF}$ term indicates a collapse of the magnetic order coincident in pressure with the structural phase transition observed in the PXRD (43.7(4) GPa) and FTIR (45.5(4) GPa), predicted by the DFT calculations (~40 GPa), and inferred from the aggregate of the XES data and the $\Delta E_Q$ vales from the SMS experiments (45.6(4) GPa).

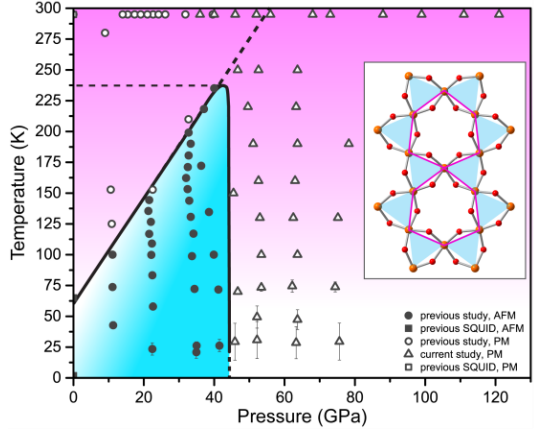

**FIG. 3** | The temperature-pressure magnetic phase diagram for jarosite up to 121 GPa. Closed symbols denote antiferromagnetic order. Open symbols denote measurements in which no magnetic ordering is observed. The antiferromagnetic region is highlighted in blue while the paramagnetic region is highlighted in pink. In the legend, AFM denotes antiferromagneic ordering, while PM denotes paramagnetic. The measured values from a previous study are reported in reference [42]. The dashed lines indicate the linear trajectory of $T_N$ with pressure and the maximum measured $T_N$ value. In the inset, blue equilateral triangles and distorted pink hexagons highlight the twisted kagomé lattice of $Fe^{3+}$ ions at 80 GPa in the $R\bar{3}c$ phase. Iron and oxygen ions are depicted as orange and red spheres, respectively, while the other atoms are ommited for clarity.

The observed magnetic collapse indicates either the addition or loss of an exchange interaction, and/or a drastic change in the existing exchange interactions. At ambient pressure, there are two principal symmetric exchange couplings, $J_1$ (near neighbor) and $J_2$ (next near neighbor), and two components of the antisymmetric exchange interaction $\mathbf{D}_{ij}$, the out-of-plane $\mathbf{D}_z$ and the in-plane $\mathbf{D}_\rho$ (i and j are site indices). An analysis of the spin wave dispersions (in ref. [17]) at ambient pressure shows that (in meV) $J_1 = 3.18$, $J_2 = 0.11$, $|\mathbf{D}_\rho| = 0.197$, and $\mathbf{D}_z = -0.196$. $J_1$ is determined by the Fe–O–Fe pathway. $J_2$ is determined by an Fe–O–O–Fe super-superexchange pathway, which explains why it is relatively weak compared to $J_1$. $\mathbf{D}$ is constrained to lie within the mirror plane bisecting the two iron atoms and thus normal to the Fe–Fe contact that comprises the equilateral triangles of the lattice [19,20].

The transition from the $R\bar{3}m$ phase to the $R\bar{3}c$ structure modifies the interatomic distances and angles and therefore affects these exchange interactions (Fig. 3, inset). The bond angle in the Fe–O–Fe pathway does not change drastically with pressure through the phase transition (Fig. S32). However, the planar O ion moves off center between the two $Fe^{3+}$ ions, creating inequivalent bond distances $d(Fe-O)_1$ and $d(Fe-O)_2$. Despite this, the equilateral triangles of $Fe^{3+}$ ions remain intact as each $Fe^{3+}$ ion has two $d(Fe-O)_1$ and two $d(Fe-O)_2$ distances. Thus, we do not expect $J_1$ to change drastically across $P^{*,2}$. Conversely, the super-superexchange, $J_2$, splits into $J_{2a}$ and $J_{2b}$. However, the changes in bond lengths and angles are such that we expect both to remain small compared to $J_1$. As with the ambient pressure phase, we expect that the interplane couplings remain negligible in the $R\bar{3}c$ structure.

The loss of the mirror plane between the $Fe^{3+}$ ions removes a symmetry constraint for the direction of $\mathbf{D}$ [20] which could influence the magnetic order. In the ambient-pressure $R\bar{3}m$ structure, $\mathbf{D}$ is constrained to lie within the mirror plane, with a negative $\mathbf{D}_z$ stabilizing a positive vector chirality and $\mathbf{D}_\rho$ causing canting of the spins out of the plane [12,17]. When the mirror plane is removed in the $R\bar{3}c$ structure, the symmetry constraint lifts and $\mathbf{D}_\rho$ can rotate, although we anticipate that this rotation is small. Moreover, the positive vector chirality state is still allowed in the $R\bar{3}c$ space group [70].

Instead, we hypothesize that a pressure-driven quenching of the antisymmetric exchange produces the observed vanishing of magnetic order. At ambient pressure, $\mathbf{D}$ lifts the energy of the kagomé zero modes which in turn promotes long range magnetic order and explains the relatively high $T_N$ [14,17]. In the $R\bar{3}c$ structure, the apical oxygen ions shift and the $(Fe-O)_{eq}$ bond splits lifting any remnant degeneracy in the crystal field splittings. This acts to quench any orbital angular momentum, $L$, arising from mixing with low-lying excited states. As $\mathbf{D}$ is proportional to $L$, a quenching of $L$ at high pressures would quench the DM term and effectively lower the zero modes back towards zero energy. The same arguments apply to the single-ion anisotropy which is an alternate source of zero mode lifting [14,17]. This could be investigated in the future by high temperature susceptibility measurements and Fe L-edge [71] spectroscopy at high pressure to test for a deviation of the Fe ions from the $S=5/2$, $L=0$ state. We note that the material may be magnetically ordered at temperatures below the limit of detection of these experiments, suggesting future investigations into the magnetic structure at these extreme conditions.

Based on the data herein, we exclude certain possible magnetic ground states in the measured region. We exclude as a possibility a spin glass state, which shows signatures in Mössbauer measurements that are not observed herein [72]. Likewise, we exclude a conventional quantum spin liquid state or a valence bond solid state given the classical nature of $S=5/2$ spins [1–4,73,74]. From the vanished $B_{HF}$, we conclude that the pressure-induced phase is distinct from the field-induced phase in jarosite [12]. Lastly, a non-zero value of the Coulomb repulsion $U$ is required to maintain a high spin state at high pressure in the DFT calculations, and as such the material is predicted to remain insulating across $P^{*,2}$ (Fig. S33). The results of the DFT calculations imply that an insulator-to-metal Mott transition does not occur. Additionally, we do not observe any optical signatures of metallization with pressure [75,76]. However, the observed linear increase in $T_N$ followed by a dramatic quenching of magnetic order is similar to the pressure-induced metallization events which occur in several transition metal halides [77–79] and we anticipate that future studies of the resistivity of this material at extreme conditions will be extremely illuminating.

Our unexpected finding of the dramatic collapse of the magnetically ordered q=0 state in jarosite at high pressure suggests there is much remaining to be discovered in the phase space of 2D magnetically frustrated materials. Additional theoretical and experimental work on the rare twisted kagomé lattice with competing symmetric and antisymmetric exchange interactions will contribute to our understanding of this unusual phenomenon.

We thank Drs. S. N. Tkachev, D. Zhang, and C. Kenney-Benson for technical support. We thank Dr. E. C. Thompson for wonderful discussions. Creating and understanding magnetic frustration in synthetically derived materials was supported by the ARO (W911NF1810006). Initial work to develop high pressure methodologies was supported by the AFOSR (FA9550-17-1-0247). S.D.J. acknowledges support from the Capital/DOE Alliance Center (CDAC) for beamtime at HPCAT, the NSF (EAR-1452344), and the David and Lucile Packard Foundation. D.P. and J.M.R. acknowledge the Army Research Office under Grant No. W911NF-15-1-0017 for financial support and the DOD-HPCMP for computational resources. W.B. was partially supported by the Consortium for Materials Properties Research in Earth Sciences (COMPRES). M.R.N. is supported by the Materials Sciences and Engineering Division, Basic Energy Sciences, Office of Science, US DOE. Part of this work was performed under the auspices of the U.S. Department of Energy by Lawrence Livermore National Security, LLC, under Contract DE-AC52-07NA27344. The gas loading was also partially supported by COMPRES under the NSF Cooperative Agreement EAR 1634415. HPCAT operations are supported by DOE-NNSA's Office of Experimental Sciences. The National Science Foundation – Earth Sciences (EAR 1634415) and Department of Energy – Geosciences (DE-FG02-94ER14466) support GSECARS. Experiments at Sector 13-BM-C were conducted under the Partnership for Extreme Crystallography (PX^2), which is also supported by COMPRES. The APS is a U.S. Department of Energy (DOE) Office of Science User Facility operated for the DOE Office of Science by ANL under Contract No. DE-AC02-06CH11357.

Supplemental Materials for:

# Pressure Induced Collapse of Magnetic Order in Jarosite

Ryan A. Klein,[1] James P. S. Walsh,[1] Samantha M. Clarke,[2] Zhenxian Liu,[3] E. Ercan Alp,[4] Wenli Bi,[5] Yue Meng,[6] Alison B. Altman,[1] Paul Chow,[6] Yuming Xiao,[6] M. R. Norman,[7,†] James M. Rondinelli,[8,†] Steven D. Jacobsen,[9,†] Danilo Puggioni,[8,†] and Danna E. Freedman[1,†]

[1] *Department of Chemistry, Northwestern University, Evanston, Illinois 60208, United States*
[2] *Lawrence Livermore National Laboratory, 7000 East Ave, Livermore, California 94550, United States*
[3] *Department of Physics, University of Illinois at Chicago, Chicago, Illinois 60607, United States*
[4] *Advanced Photon Source, Argonne National Laboratory, 9700 South Cass Avenue, Lemont, Illinois 60439, United States*
[5] *Department of Physics, University of Alabama at Birmingham, Birmingham, Alabama, 35294, United States*
[6] *HPCAT, X-Ray Science Division, Argonne National Laboratory, Lemont, Illinois 60439, United States*
[7] *Materials Science Division, Argonne National Laboratory, Argonne, Illinois 60439, United States*
[8] *Department of Materials Science and Engineering, Northwestern University, Evanston, Illinois 60208, United States*
[9] *Department of Earth and Planetary Sciences, Northwestern University, Evanston, Illinois 60208, United States*

# Table of Contents

**Full experimental details**

**Synthesis of jarosite, $KFe_3(OH)_6(SO_4)_2$:** We synthesized jarosite according to previously published methods.[1–3] Both the non-enriched and 20% $^{57}Fe$ enriched jarosite are synthesized as in Klein, *et al.*, Reference 1.

**Details of DAC assembly:** We employed symmetric Princeton-type diamond anvil cells (DAC) during all of the experiments enumerated below. We used Type Ia (100)-oriented diamonds with culets of either 200 or 300 µm. For the megabar experiments, we used beveled diamonds with 100 µm culets and 300 µm bevels. For the diffraction experiments, we used anvils of the Boehler–Almax design fitted to 80° conical opening tungsten carbide seats. We used pre-indented rhenium gaskets (3.3 × 3.3 × 0.25 mm squares). We used either a micro-electrical discharge machine or the laser drilling apparatus at HPCAT, the Advanced Photon Source (APS), Argonne National Laboratory (ANL) to drill cylindrical holes of 50–170 µm in diameter in the center of the indents in the gaskets.[4] We loaded small ruby spheres and/or small chips of platinum alongside the samples to act as pressure calibrants.[5] The cells were gas loaded with neon as the pressure-transmitting medium.[6] Potassium bromide[7] or magnesium oxide[8] was used in cases where neon was not the pressure-transmitting medium. Sample pressure was monitored during the initial compression using the ruby $R_1$ fluorescence peak.[9] The intensity of the ruby fluorescence signal decreases with increasing pressure, and ruby fluorescence spectra can be challenging to measure in a cryostat. Therefore, when ruby fluorescence was not used at the pressure calibrant, either the neon or platinum equations of state or the diamond anvil Raman signal were used to infer the measurement pressure.[10–12]

**Details of powder x-ray diffraction:** We conducted *in situ* synchrotron powder x-ray diffraction (PXRD) experiments at beamlines 16-ID-B and 16-BM-D, HPCAT, APS, ANL.

For the ambient-temperature equation of state data set, we employed a DAC with 200 µm culet diamonds and a rhenium gasket with a drilled hole of about 100 µm in diameter and 35 µm pre-indentation thickness. The sample space contained a small ruby sphere and several very small jarosite crystallites. The cell contained neon, which acted as the pressure-transmitting medium and as the pressure calibrant.[6,11] A micro-focused beam of synchrotron x-ray radiation irradiated the sample. The beam was focused to 3 µm × 6 µm at FWHM with an incident wavelength of λ = 0.406600 Å.

For the variable-temperature, variable-pressure equation of state data sets, we employed a DAC with 200 µm culet diamonds and a rhenium gasket with a drilled hole of about 100 µm. We ground together fine platinum powder with jarosite using a mortar and pestle and loaded this powder in the DAC. This DAC was gas loaded with Ne gas which acted as the pressure-transmitting medium. The experiment used a double-diaphragm gas membrane to control the pressure, and a helium cryostat with a heater to control the temperature. The pressure was inferred from the measured Pt unit cell volume using the Pt equation of state.[8] The temperature was measured using two thermocouples. The first thermocouple was placed between the copper block cooled by the He off-gassing and the DAC. The second thermocouple was placed in the window of the DAC, close to the sample, and was put into contact with the DAC using thermally conducting grease. The reported temperatures are the average of the measured temperatures from the thermocouples. The reported error in temperature is the half difference of the measured temperatures. The beam was focused to 3 µm × 6 µm at FWHM with an incident wavelength of $\lambda = 0.413300$ Å.

In both diffraction experiments, we collected diffraction images as the cell was rotated around $\Omega$ (rotation axis perpendicular to the x-ray beam) over the range −10° to 10°. At each pressure, we measured a 3 × 3 grid in the plane normal to the incident x-rays using a grid spacing of 5 µm. We used a double-diaphragm gas membrane to control the pressure. We collected diffraction images using a MAR CCD detector, and we used the Dioptas 0.3.2.beta software package to integrate the images to produce the corresponding 1D diffraction patterns.[13]

**Analysis of the powder x-ray diffraction patterns:** We analyzed the powder diffraction patterns using the Bruker AXS software package TOPAS, version 5.[14] We fit the background to an 8-parameter Chebyshev polynomial and pseudo-Voigt line broadening was employed for all phases. We fit the patterns over the $2\theta$ range 3° to 22° to extract unit cell parameters of all phases at each pressure. We used the Pawley method to extract the unit cell parameters of all relevant phases at every condition measured. The powder patterns can be Pawley fit to both the $R\bar{3}m$ and the $R\bar{3}c$ space groups throughout the pressure range for every experiment. The fit statistics in each case are very similar because the hkl positions the models are nearly identical (Fig. S7). The unit cell parameters and the pressure for each pattern collected are tabulated in Tables S1, S2 for the ambient-temperature equation of state data set, and in Tables S3 for the variable temperature, variable pressure data set. The patterns are plotted in Fig. S3 and Fig. S11 for the ambient temperature and the variable temperature data sets, respectively.

**Details on the equation of state fitting:** We used the software package EoSFit-7c to fit the unit cell parameters as a function of pressure (Fig. S4–S6).[15] We used a third-order Birch–Murnaghan (BM3) equation of state to fit these data (Equation 1).[16,17] The parameters in the BM3 equation are: $P$ is pressure, $V$ is volume, $V_0$ is the volume at zero applied pressure, $B_0$ is the bulk modulus, and $B_0'$ is the first derivative of the bulk modulus with respect to pressure.

$$P(V) = \frac{3B_0}{2}\left(\left(\frac{V_0}{V}\right)^{\frac{7}{3}} - \left(\frac{V_0}{V}\right)^{\frac{5}{3}}\right)\left\{1 + \frac{3}{4}(B_0' - 4)\left(\left(\frac{V_0}{V}\right)^{\frac{2}{3}} - 1\right)\right\} \quad \text{(Eq. 1)}$$

Two discontinuities manifest between ambient pressure and 78.6(7) GPa, such that a single BM3 is inadequate for fitting the entire pressure range. We therefore split the data into three groups: a low-pressure region (7.9(1)–15.9(1) GPa), a mid-pressure region (19.5(1)–39.4(3) GPa), and a high-pressure region (49.9(5)–78.6(7) GPa). For comparison, we also modeled the entire pressure range (7.9(1)–78.6(7) GPa) by itself. For each pressure region, we used a BM3 to model both the volume and the axial compressibility in the $a$-axis and the $c$-axis. We evaluated the axial compressibility by cubing the $a$- and $c$-axis lengths, respectively, and fitted them using BM3 equations, resulting in a linearized modulus, $M$. The parameters for the 12 resulting fits are given in Table S4. We weighted all refinements against errors in both pressure and in the unit cell parameters; the values and their errors are given in Tables S1, S2.

**Details of the Fourier-transform infrared spectroscopy:** All FTIR experiments were conducted at the NSLS-II at beamline 22-IR-1. We prepared two diamond anvil cells for this experiment, the first with 300 μm culet diamonds and the second with 200 μm culet diamonds. We loaded the first cell with a thin single crystal of jarosite, a ruby sphere, and neon. We measured this cell up to ~38 GPa. We loaded the second cell with KBr, a ruby sphere, and a thin, pressed flake of crystalline jarosite. We measured this cell up to ~65 GPa. Once the ruby bridged the diamonds in this experiment, between 30 and 35 GPa, we used the diamond first-order Raman edge to measure the pressure in the cell.[18] We allowed the cell to sit for several days at ~65 GPa, then increased the pressure to ~70.0(7) GPa and measured the FTIR spectra for the sample upon decompression. There is minimal hysteresis in the transitions in the spectra and the transitions are fully reversible with pressure. The data are plotted in Fig. S12–S14.

**Analysis of the Fourier-transform infrared spectra:** We fit the background-corrected spectra using a fifth order polylogarithmic curve for the background and using Lorentzian curves to fit the features. By fitting the features in the spectra with Lorentzian curves, we extracted the peak position and FWHM value for each mode at each pressure. For the data that originated from the lower-pressure cell, the background was corrected after the data collection. The background used came from the same two anvils with exactly the same optical configuration such as the aperture size. For the second cell, the background was obtained at each pressure, from an area in the DAC without sample present. The second cell was used to collect data upon compression and upon decompression. The mode positions as a function of pressure are plotted in Fig. S15. An analysis of the FWHM values for the modes in the high-wavenumber region is shown in Fig. S16, while an analysis of the FWHM values for the modes in the low-wavenumber region is shown in Fig. S17.

**Details of synchrotron Mössbauer spectroscopy:** We conducted three separate sets of SMS experiments. For all three sets of experiments, we used jarosite enriched 20% in $^{57}$Fe. The data collected in these experiments are plotted in Figures S22–S27, and the fit parameters, temperatures, and pressures are given in Tables S5, S6.

For the ambient temperature experiments conducted while the synchrotron operated in hybrid bunch mode (Fig. S22), we collected the data at Sector 3, APS, ANL. We used a DAC with 200 μm culet diamonds. We used a pre-indented rhenium gasket and loaded the DAC with a single crystal of enriched jarosite, a ruby, and neon. The use of the synchrotron's hybrid bunch mode allowed us to resolve features in the spectra out to 400 ns. We measured the pressure in the DAC using an off-line ruby fluorescence system to measure the ruby $R_1$ line. The raw data and the corresponding fits are plotted in Fig. S22, and the fit parameters, including the $\chi^2$ values, are listed in Table S5.

For the ambient temperature experiments up to 121(7) GPa (Fig. S23), we collected the data at beamline 16-ID-D, HPCAT, APS, ANL while the synchrotron operated in the standard 24 bunch mode. We used a DAC with diamonds with 100 μm culets beveled out to 300 μm. We loaded the DAC with a single crystal of enriched jarosite and we used MgO as the pressure-transmitting medium. We measured the pressure using the diamond anvil Raman signal.[12]

For the variable temperature experiments, we collected the data at beamline 16-ID-D, HPCAT, APS, ANL while the synchrotron operated in the standard 24 bunch mode (Figs. S23–S27). We used a DAC with 200 μm culet diamonds. We loaded the DAC with a single crystal of enriched jarosite and a small ruby sphere, and we used neon as the pressure-transmitting medium. We measured the pressure using the ruby $R_1$ fluorescence feature.[5] We controlled the pressure using a double-diaphragm membrane and we controlled the temperature using a cryostat with a heater. We measured the temperature using two thermocouples as described above for the variable temperature PXRD experiments. We report the measured temperature in the same way for the SMS and the PXRD data.

**Analysis of the synchrotron Mössbauer spectra:** For the three data sets, we fit the spectra using CONUSS-2.1.0 (W. Sturhahn, www.nrixs.com). We allowed the following parameters to freely refine: $\Delta E_Q$, α, β, sample thickness, and the scaling factor. Please see the detailed manual for CONUSS for a description of each of these parameters. We set the energy/time resolution to 1 ns and then allowed it to refine near that value. We set the texture to 100% because the samples were all single crystals. We fixed the Lamb-Mossbauer factor at 0.796, and the abundance of the Mössbauer atom at 20%.[19] The relevant fit parameters, including the $\chi^2$ values, are listed in Table S4. We did not observe magnetic ordering in any of the data sets, therefore $B_{HF}$ was set to zero and not refined.

For the variable temperature set of experiments, we used a standard binning and fit the spectra from 25 to 130 ns. The data were modeled well with one site at every condition measured. The fit parameters are listed in Table S5.

For the hybrid mode experiments, we used a binning of four, instead of the standard ten, for our data. We fit the spectra from 25 to 400 ns. We fit the spectra using one iron site in all but two cases. At 46.0(4) GPa, we fit the data with two nearly identical sites that differed in their quadrupole splitting value. The parameters for both sites are listed in Table S6, with the second site denoted by an asterisk (*). At 73.0(7) GPa, we fit the data with two sites to approximate a strain-induced distribution in both the quadrupole splitting and the isomer shift. The second site is denoted by an asterisk (*).

For the megabar set of experiments, we used a standard binning and fit the spectra from 25 to 130 ns. The data were modeled well with one set except at 121(7) GPa, in which a minority site was used to approximate strain in the relevant fit parameters, as above. The fit parameters are listed in Table S6.

**Details of x-ray emission spectroscopy:** We collected the XES data at ambient temperature at 16-ID-D, HPCAT, APS, Argonne National Laboratory. We indented and drilled the beryllium gaskets, and loaded the cells, in the HPCAT wet lab adjacent to the beam line. We drilled a hole in the beryllium gasket with a diameter of ~100 μm. We compressed this gasket until it was ~15-30 μm thick, and then redrilled the hole to ~ 80 μm. We loaded the cell with a thin single crystal of jarosite and a small ruby sphere, and KBr as the pressure transmitting medium. We monitored the $R_1$ ruby fluorescence feature to measure the pressure for each spectrum. We collected spectra between 1.1(1) GPa and 75.0(7) GPa. We recommend using an insert of cBN with epoxy when attempting low-temperature, high-pressure measurements using a Be gasket.

**Analysis of the x-ray emission spectra:** We normalized the raw data with respect to the beam flux at the time of each data acquisition. Then, we truncated the data to an energy range of 7020 and 7080 eV. Next, we performed a baseline subtraction and normalized each spectrum such that the area underneath the curve between 7020 and 7080 eV became equal to one by integrating each spectrum and divided each data point by the resulting integral value. The difference curves are equal to the 1.1 GPa spectrum minus the higher-pressure spectra. We integrated the absolute value of these difference curves to give the corresponding IAD value.[20] The error bars in the IAD value are equal to 3σ for the absolute value of the difference curves.

**Calculation Details:** We perform spin-polarized density functional calculations within the Perdew-Burke-Ernzerhof exchange-correlation functional revised for solids (PBEsol)[21] and the PBE + $U$ method[22] as implemented in the Vienna *Ab initio* Simulation Package (VASP)[23] with the projector augmented wave (PAW) method[24] to treat the core and valence electrons using the following electronic configurations: $3d^74s^1$(Fe), $3s^23p^64s^1$(K), $3s^23p^4$(S), $2s^22p^4$ (O), and $1s^1$ (H). To avoid spin crossover as a function of pressure we use a $U$ value of 5 eV. A kinetic energy cutoff energy of 650 eV is used to expand the wave functions and a centered $6 \times 6 \times 6$ k-point mesh combined with the tetrahedron and Gaussian methods are used for Brillouin zone integrations. The ions are relaxed toward equilibrium until the Hellmann-Feynman forces are less than 1 meV Å$^{-1}$, whereas the cell parameters are fixed to the experimental values. *Frozen-phonon* technique with a 2x2x2 supercell, ferromagnetic ordering, and the phonopy package[25] are used for the phonon calculations.

**Details of the high-pressure Rietveld refinements and structure solution:** To determine the structure in the high-pressure phase of jarosite, we conducted PXRD and FTIR experiments, and DFT calculations. The aggregate of the results was used to conclude that the ambient-pressure $R\bar{3}m$ structure transforms into the $R\bar{3}c$ structure at pressures above $P^{*,2}$=43.7(4) GPa. The analysis includes an examination of the Pawley fits of the PXRD patterns above the phase transition pressure, the FTIR spectra below and above the phase transition pressure, DFT calculations of the energy for possible new structures above the phase transition pressure, and Rietveld refinements of a representative high-pressure PXRD pattern using the possible models identified by the Pawley analysis, the FTIR spectra, and the DFT calculations. This analysis is detailed below.

The PXRD patterns do not display any additional Bragg peaks across the phase transition pressure at $P^{*,2}$ within the limit of resolution of the synchrotron experiments. From the lack of any additional resolved Bragg peaks in the patterns at high pressure, we conclude that the Bravais lattice likely remains constant across the phase transition. As such, we limit the number of reasonable space groups for jarosite's high-pressure phase to the trigonal space groups with a group-subgroup or group-supergroup relationship to $R\bar{3}m$, namely, $R\bar{3}m$, $R\bar{3}c$, $R\bar{3}$, $R3m$, and $R32$. We performed a thorough Pawley analysis of the PXRD patterns above the phase transition pressure using these five space groups. The space groups yield identical (or practically identical, see Fig. S7 as an example) sets of Miller indices such that the Pawley fit alone cannot rule out any of these five possible space groups.

The FTIR experiments definitively show a lowering in symmetry across the phase transition at ~45 GPa. The number of observed IR modes nearly doubles across the transition in the measured spectral window (Figs. S20−S22). This observation rules out the ambient pressure structure as a reasonable model for the high-pressure phase. In other words, above 40 GPa, the increase in the number of observed modes illustrates the appearance of new a structural phase, different from the original $R\bar{3}m$ structure.

Using DFT calculations, we found two pressure-induced lattice instabilities located at the Γ and T (0,0,3/2) points of the Brillouin zone, respectively. Upon condensing these instabilities and after atomic relaxation, we were able to stabilize two high-pressure phases with $R\bar{3}$ and $R\bar{3}c$ space groups, respectively, with the latter being lower in energy. These calculations limited the number of possible space groups which can describe jarosite's high-pressure phase from 5 to 2.

We then selected a representative powder X-ray diffraction pattern at a pressure well above the phase transition and performed Rietveld refinements using the parent $R\bar{3}m$ structure and the two calculated structures with $R\bar{3}$ and $R\bar{3}c$ symmetry as starting models for this additional analysis. The $R_{exp}$ value for the selected 62.1(6) GPa pattern is $R_{exp} \approx 2.2$, which is typical of the high-pressure patterns and of high-pressure patterns collected at these conditions in general. The relatively high $R_{exp}$ value arises from multiple intrinsic experimental factors, including pressure-induced line broadening and a curved background originating from the diamond anvil cell itself. Additionally, an examination of the 2D detector images shows relatively poor powder averaging for the sample (see Fig. S2). The poor powder averaging arises from the nature of the sample and experimental requirements of the DAC. In every synthesized batch, the sample comprises single crystals of varying size, the smallest of which are on the order of $\sim 5 \times 5 \times 5$ μm$^3$. The sample chamber is a cylinder ~35 μm in height and ~100 μm in diameter at ambient pressure and shrinks appreciably as pressure increases. To avoid bridging the sample space at high pressure, and thus compromising the quasi-hydrostatic conditions of the experiment, we took great care to avoid overloading the sample space of the DAC. This need is balanced with the need to sample enough orientations of the single crystals to yield a useable powder average during the experiment. To increase the number of crystalline orientations sampled, a $3 \times 3$ grid in real space was measured using a step size slightly larger than the FWHM of the X-ray beam. The 9 resulting patterns are averaged. The pressure gradient over this space also contributes to line broadening in the PXRD patterns. Thus, the single-crystalline nature of the sample at the size scales required to conduct this experiment yield a poor powder averaging, necessitating the use of preferred orientation terms in the Rietveld refinement. There is currently no pathway to collecting significantly higher quality data (with a lower $R_{exp}$ value or less preferred orientation effects).

In all three refinements, isotropic thermal parameters of $b_{eq}=0.5$ were used for every atom. The occupancy was fixed at unity for all atoms. The background was fitted using a five-term shifted Chebyshev polynomial function, which was initialized using the same values for all five terms across the three refinements and then allowed to vary. The peak profile was fitted during the Pawley fits using Lorentzian size and strain terms only, and these terms were used and held constant for all three Rietveld refinements. The scale of each phase was refined freely. The unit cell parameters were initialized as those found during the Pawley fitting and then refined. The refineable atomic positions were fixed to be equal to the positions from the DFT calculations for the $R\bar{3}$ and $R\bar{3}c$ structures and fixed to be equal to the positions from the ambient pressure structure for the $R\bar{3}m$ model. As a note, there are only 8 general coordinates (out of a total of 18) which are held fixed in each refinement (See Table S7). A spherical harmonics model with four terms (the minimum possible number of terms achievable in Topas Academic) was used to model the preferred orientation effects for all three Rietveld refinements. The results of the Rietveld refinements are summarized in Table S7.

The Rietveld refinements using the three different models yielded fits with fit statics of $R_{wp}<R_{exp}$ and $R_p<R_{exp}$, even when the atomic positions were not refined. As such, Rietveld refinements of the powder diffraction data are unable to distinguish between these three models, which all fitted the data equally well.[26] Importantly, the Rietveld analysis shows that the calculated $R\bar{3}c$ phase accurately models the data.

Lastly, note that while the calculated $R\bar{3}m$ and $R\bar{3}$ structures are described by one formula unit, the $R\bar{3}c$ phase is described by two formula units (*via* a doubling along the *c*-axis). In Fig. S22, for the high-wavenumber region, we compare the calculated zone-center (Γ-point) IR modes of the $R\bar{3}$ phase with the experimentally observed ones. The main result is that the $R\bar{3}$ phase, as with the parent $R\bar{3}m$ phase, cannot justify the doubling of the observable IR modes in the high frequency region. Conversely, at high pressure, we calculated four IR modes for the $R\bar{3}c$ phase, in good qualitative agreement with the experimental observation.

For the above reasons, we deduce that the $R\bar{3}c$ model correctly and accurately describes jarosite's high-pressure phase and we conclude that the high-pressure phase contains a twisted kagomé lattice.

**Interpretation of the Fourier-transform infrared spectra:** The spectral changes in the FTIR data are summarized in Figs. S15 and S16. The evolution of the *v*(OH) modes with pressure illustrates the overall behavior of the sample under compression. To facilitate discussion, we divide the data into four pressure intervals corresponding to the highlighted regions (Fig. S15) and discuss the evollution of the *v*(OH) modes followed by a discussion of the low-wavenumber region of the spectra.

At ambient pressure, there is one very sharp *v*(OH) symmetric stretching frequency in jarosite at 3387.3 cm$^{-1}$ with a broad shoulder feature [2]. These both redshift with pressure smoothly and monotonically in the highlighted region 1 (Fig. S15). In this region, jarosite hosts relatively weak hydrogen bonding. In region 2, the modes continue to blueshift and both *v*(OH) modes stiffen; particularly the higher energy mode. The onset of region 2 coincides in pressure with the isomorphous phase transition in the PXRD data at $P^{*}$,[1]. This phase transition in part arises from a change in the compressability in the *c*-axis as evidenced by the plot of the $c/a$ ratio. This change in compressability originates from the stiffening of the *v*(OH) modes which signals a strengthening of the hydrogen bond network.[1] In the third highlighted region in Fig. S15, the modes flatten and become pressure independent. The *v*(OH) frequency suggests strong, but not necessarily symmetric, hydrogen bonds.[27–32] The pressure-independent nature of the modes signals that the hydrogen bond network is quite strong, in agreement with the large axial bulk modulus, $M_0$, in the $c$-axis direction. At $P^{*,2}$ there is a large discontinuity in all modes which defines the pressure boundary between the third and fourth highlighted regions in the figure. Above this pressure, two additional modes appear at high wavenumber, and the two existing modes shift in frequency, based on the analysis of the FWHM values with pressure of the various modes (Fig. S16).

The pressure intervals defined as regions 1–4 in Fig. S15 correspond with spectral changes in the low-wavenumber region of the spectra (Fig. S16). At ambient pressure, the low-wavenumber plot is dominated by sulfate stretching modes, including higher frequency antisymmetric sulfate stretching modes [$\nu_3(SO_4^{2-})$], symmetric sulfate stretching mode [$\nu_1(SO_4^{2-})$], and out-of-plane bending sulfate modes [$\nu_4(SO_4^{2-})$] at the lowest frequencies.[33–35] These modes are labeled in the figures. The weak $\delta(OH)$ deformation mode at 1005 cm$^{-1}$ previously reported[2,21,22] at ambient conditions is unresolved in our data due to inherent limits in signal strength attainable inside a diamond anvil cell. The modes present in this low-wavenumber region all change as a function of pressure across the four pressure regions defined in the main text.

In region 1 the sulfate modes uniformly and monotonically blueshift. Here, jarosite is characterized by relatively weak hydrogen bonding. The hydrogen bonds in this region are non-linear and asymmetric ($\sphericalangle(O–H \cdots O) \neq 180°$).

At the low pressure boundary of region 2 a new mode emerges at intermediate frequency. We assign this mode as the $\delta(OH)$ mode due to its frequency and redshifting behavior consistent with the $\nu(OH)$ modes. We posit that the intensity of this mode may increase with pressure so as to become resolvable at this pressure. Concurrently, a new $\nu_1(SO_4^{2-})$ mode appears in the spectra from underneath an existing mode. Interestingly, one of these $\nu_1(SO_4^{2-})$ modes smoothly redshifts with pressure while the others blueshift. The onset of region 2 is marked by the initial phase transition in the PXRD data at 19.5(1) GPa, which partially arises from a change in the compressability in the $c$-axis as evidenced by the plot of the $c/a$ ratio (Fig. 1, bottom). This change in compressability coincides with a stiffening of the two $\nu(OH)$ modes, as shown by the high-wavenumber FTIR data. Concomitantly, one of the symmetric $\nu_1(SO_4^{2-})$ stretching modes, which may be related to the S–O$_{apical}$ bond, redshifts in response to the stronger O–H$\cdots$O interaction. The $c$-axis becomes increasingly less compressable as the hydrogen bonds compress. We hypothesize that the stiffening of the hydrogen bonds arises as the hydrogen bonds enter an asymmetric, possibly near-linear state ($\sphericalangle(O–H\cdots O)\approx 180°$).

In region 3, the lower of the two new $\nu_1(SO_4^{2-})$ modes and the $\delta(OH)$ mode flatten and their frequencies become pressure independent. In region 3, we hypothesize that the hydrogen bonds are strong, asymmetric, (O–H---O) and possibly near-linear ($\sphericalangle(O–H---O)\approx 180°$). The $\nu(OH)$ and $\delta(OH)$ modes are pressure-independent, which may signal that some other moiety compresses more readily in this pressure region, which is in line with the calculated rotation of the sulfate groups. Notably, the $O_{sulfate,apical}$ hosts three hydrogen bonding interactions, which makes the hydrogen bond network in jarosite extremely strong at these conditions.

A large discontinuity defines the boundary between regions 3 and 4, as the lower frequency $\nu_4(SO_4^{2-})$ mode suddenly red shifts and splits into two modes. The higher frequency $\nu_4(SO_4^{2-})$ mode rapidly blue shifts and splits into two modes. $\delta(OH)$ dies. The two new $\nu_1(SO_4^{2-})$ modes from region 2 both shift, and, lastly, the two $\nu_3(SO_4^{2-})$ diverge at the discontinuity. In region 4, all but two of the $\nu_4(SO_4^{2-})$ modes red shift with pressure. The modes at ~1050 and ~1150 cm$^{-1}$ both red shift with pressure. In region 4, the $\nu_1(SO_4^{2-})$ and $\nu_3(SO_4^{2-})$ modes blue shift, signalling that the $O_{sulfate,apical}$ atoms are less constrained in the $c$-axis direction by the hydrogen bonds, consistent with the $O_{sulfate,apical}$ forming weaker hydrogen bond interactions with the H atoms. Conversely, two sulfate modes red shift slightly, indicating that the $O_{sulfate,basal}$ atoms are more constrained.

**Notes on the Nature of the Magnetic Structure:**

The near-neighbor Heisenberg model on a kagomé lattice does not order due to the presence of zero modes (that is, rotations of the spins that cost no energy). Long range order is promoted by a non-zero next-near-neighbor interaction, $J_2$, and/or by the DM interaction (or single-ion anisotropy). A positive $J_2$ leads to q=0 order, as does the DM interaction, so they reinforce each other. Spin-wave analysis[36,37] indeed indicates that $J_2$ is positive (though small), and that the order is most likely driven by the larger DM term (though fits assuming single ion anisotropy is dominant instead work almost as well). This analysis finds that **D**$_z$ is negative, which promotes a positive vector chirality solution, with the role of **D**$_\rho$ being to determine the small canting of the spins out of the kagomé plane.

How does this change with pressure? Because **D** is proportional to $J$, and $J$ scales as $t_{pd}^4$ (where $t_{pd}$ is the Fe 3d – O 2p hopping energy), then as the Fe–O bond length decreases, one expects a dramatic rise in the magnitude of **D**, consistent with the dramatic rise in $T_N$, noting that typically, $t_{pd}$ scales as $d^{-4}$ where d is the Fe-O planar bond length. This is lessened somewhat by the decrease of the Fe–O–Fe bond angle and can be affected as well by changes in the crystal field splitting. Now we can ask what should occur once the $R\bar{3}c$ distortion sets in. First, the **D** vector no longer lies in the local mirror plane between two Fe ions. On the other hand, both the change in the tilt angle of the FeO$_6$ octahedra and the rotation of the Fe-Fe triangles is small enough that we expect only small changes in the **D** vector in regards to its magnitude and orientation (though even such small rotations of **D** could potentially impact the magnetism). In this context, the ambient pressure phase (q=0, positive vector chirality state) is an allowed magnetic ground state for the $R\bar{3}c$ space group (see Fig. S1).[38]

This suggests that we need to search elsewhere regarding the collapse of the magnetic order. First, a finite DM (or finite single ion anisotropy) is equivalent to having a finite orbital moment. This might seem somewhat unusual in the present case, given that the nominal ground state is $S = 5/2$, $L = 0$, but experimental L-edge data on iron jarosite does indeed indicate a finite $L$,[39] consistent with the fact that the high temperature moment from the Curie-Weiss susceptibility is significantly larger than the nominal value of 5.9 $m_B$. So, how can this be affected by the distortion? The strong distortion of the $FeO_6$ octahedra (especially the large predicted displacements of the apical oxygen ions) will lead to a significant splitting of the $xz$ and $yz$ orbitals, which in turn will act to quench any residual $L$. But on more general grounds, the increasing covalency of the Fe–O bonds under pressure will act to suppress magnetism, and at high enough pressures, will eventually lead to a low spin ground state and/or metallization.

One can make a simple estimate for the orientation of **D**. Take the vector that connects the midpoint of the Fe-Fe contact to the oxygen ion of that pair. For the ambient pressure phase, this leads to an orientation for **D** consistent with the spin-wave analysis, with the ratio $|\mathbf{D}_\rho|/\mathbf{D}_z$ of 1.1. Assuming the same for the calculated 80 GPa phase, one finds this ratio decreases to 0.9, with $\mathbf{D}_\rho$ rotating by ~6° relative to its $R\bar{3}m$ orientation. These deviations are significantly less for the predicted 40 GPa $R\bar{3}c$ phase.

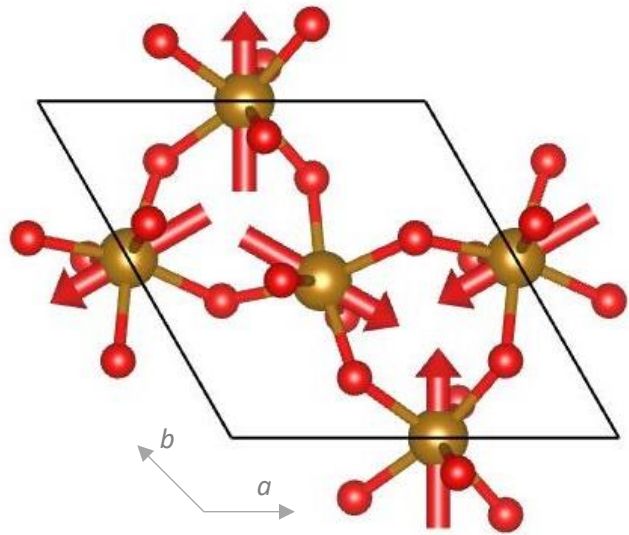

**Figure S1 |** The q=0 positive vector chirality state with ion locations given by the 80 GPa $R\bar{3}c$ prediction. This state corresponds to a magnetic space group of $R\bar{3}'c$ or $R\bar{3}c'$ depending on the magnetic stacking along $c$ (here, $'$ denotes time reversal). Only Fe (gold) and O (red) ions are shown associated with a single kagomé plane, with the arrows denoting the magnetic moments. This magnetic structure could be potentially altered by the small rotations of **D** associated with the $R\bar{3}c$ distortion, noting that the observed magnetism actually collapses in the $R\bar{3}c$ phase.

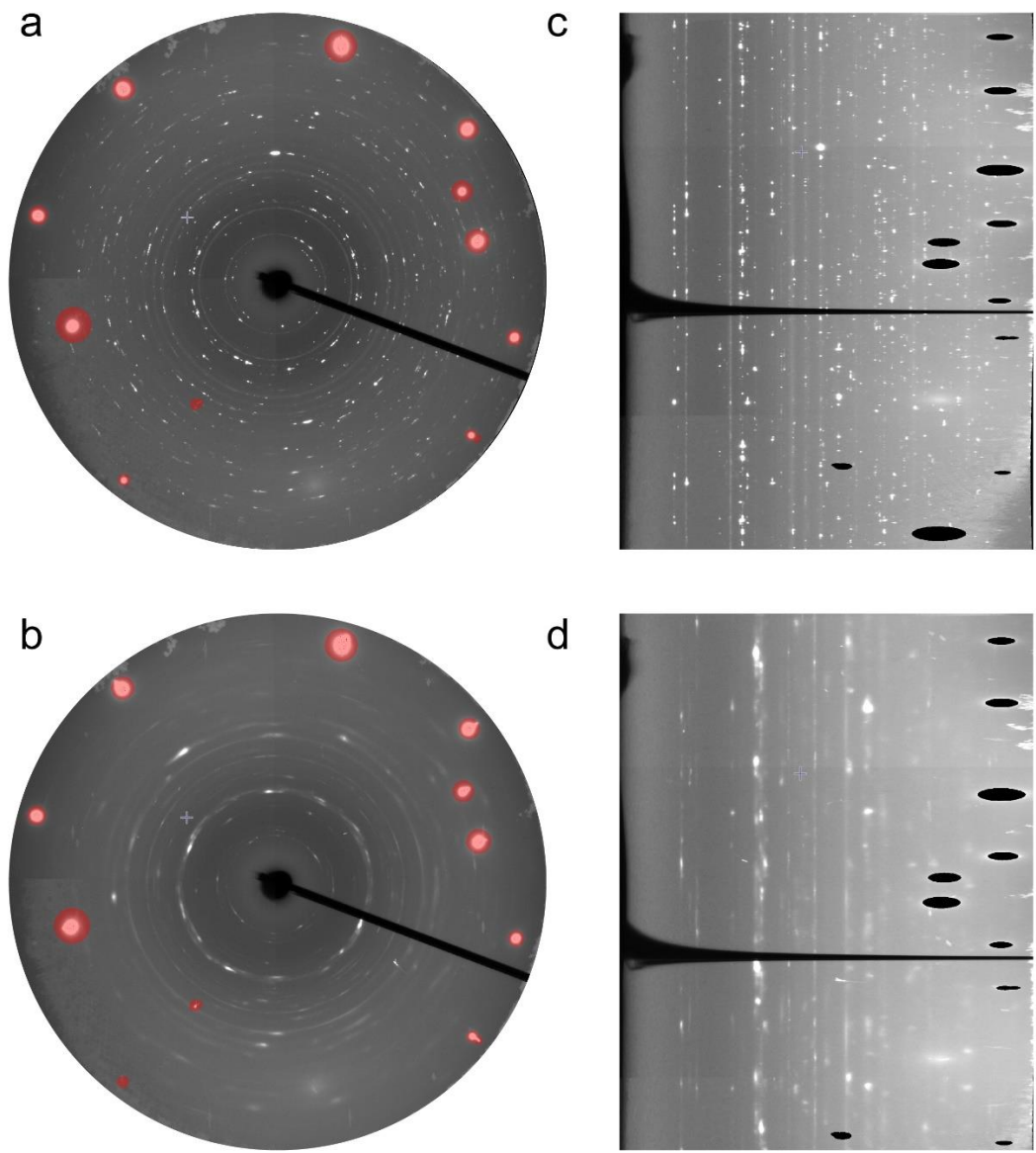

**Figure S2 |** 2D image plate images from diffraction data acquired at 7.8(1) GPa (**a**) and at 78.6(7) GPa (**b**). The images shown are the sum of nine images obtained from a 3 × 3 grid scan where the sampled spots were 5 μm apart. This summation allows for the analysis of more crystallite orientations at each pressure. **c** and **d** are the cake images that correspond to diffraction images **a** and **b**, respectively. The small, round spots are diffraction from the sample. The faint complete rings are the diffraction from platinum. The light red, transparent mask covers diamond diffraction spots. The bright, elongated and widened spots are the diffraction from the crystallized neon. We acquired these data at ambient temperature with λ = 0.406600 Å. We integrated these images to yield the 1D diffraction patterns using Dioptas.

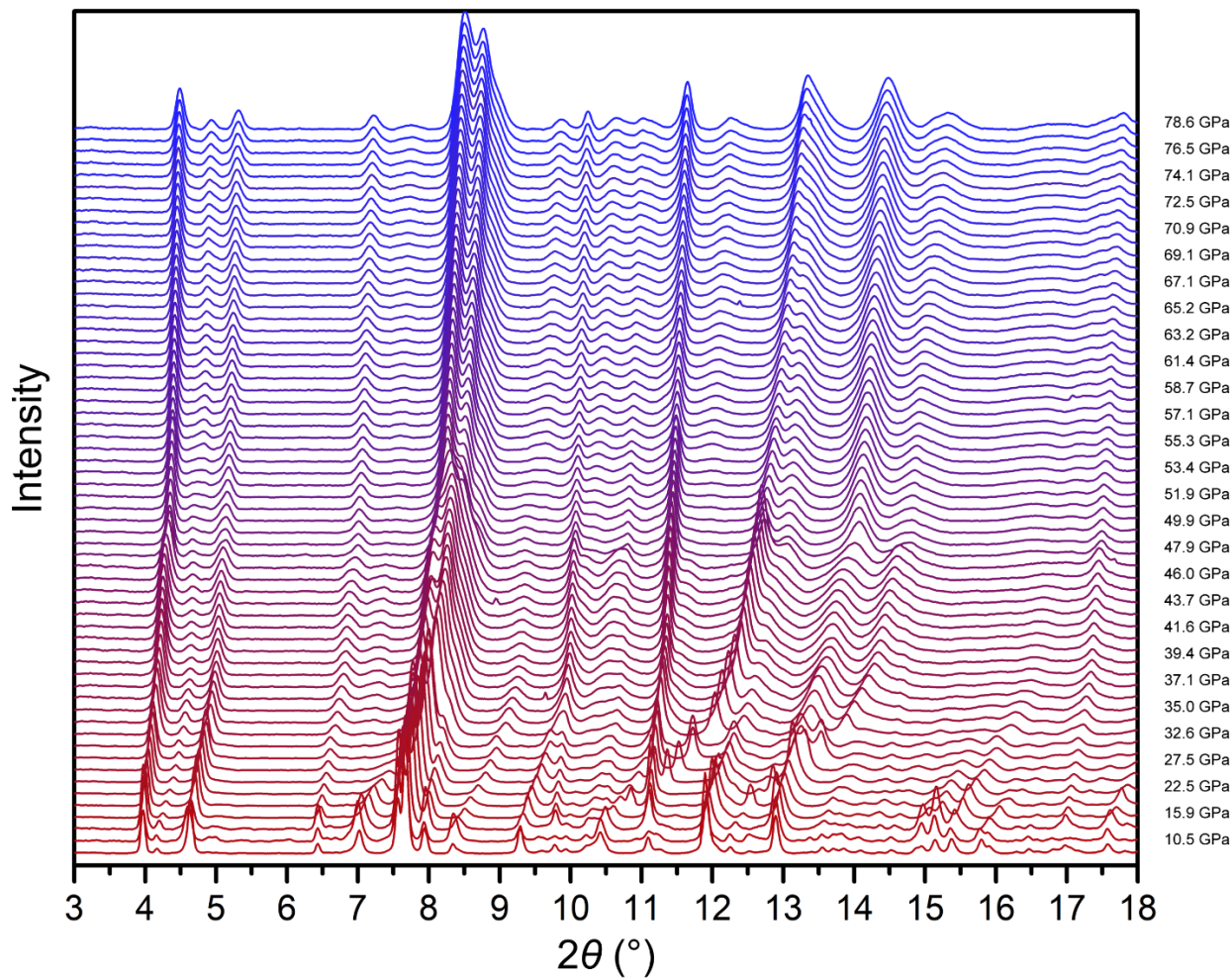

**Figure S3 |** Select PXRD patterns for jarosite as a function of pressure. Also present in these patterns are peaks from platinum and from neon. The jarosite phase is well modeled in the $R\bar{3}m$ space group at all pressures. The approximate pressure for the patterns are listed to the right of the plot. See Figures S3–S5 for a plot of the unit cell parameters, derived from these patterns, as a function of pressure. See Tables S1, S2 for the tabulated values of the unit cell parameters and the pressures for all of the collected patterns. The patterns are normalized with respect to the most intense peak, which is the intense peak between 7.5° and 8.5° at all pressures. The patterns are offset arbitrarily. We acquired these data at ambient temperature with $\lambda = 0.406600$ Å.

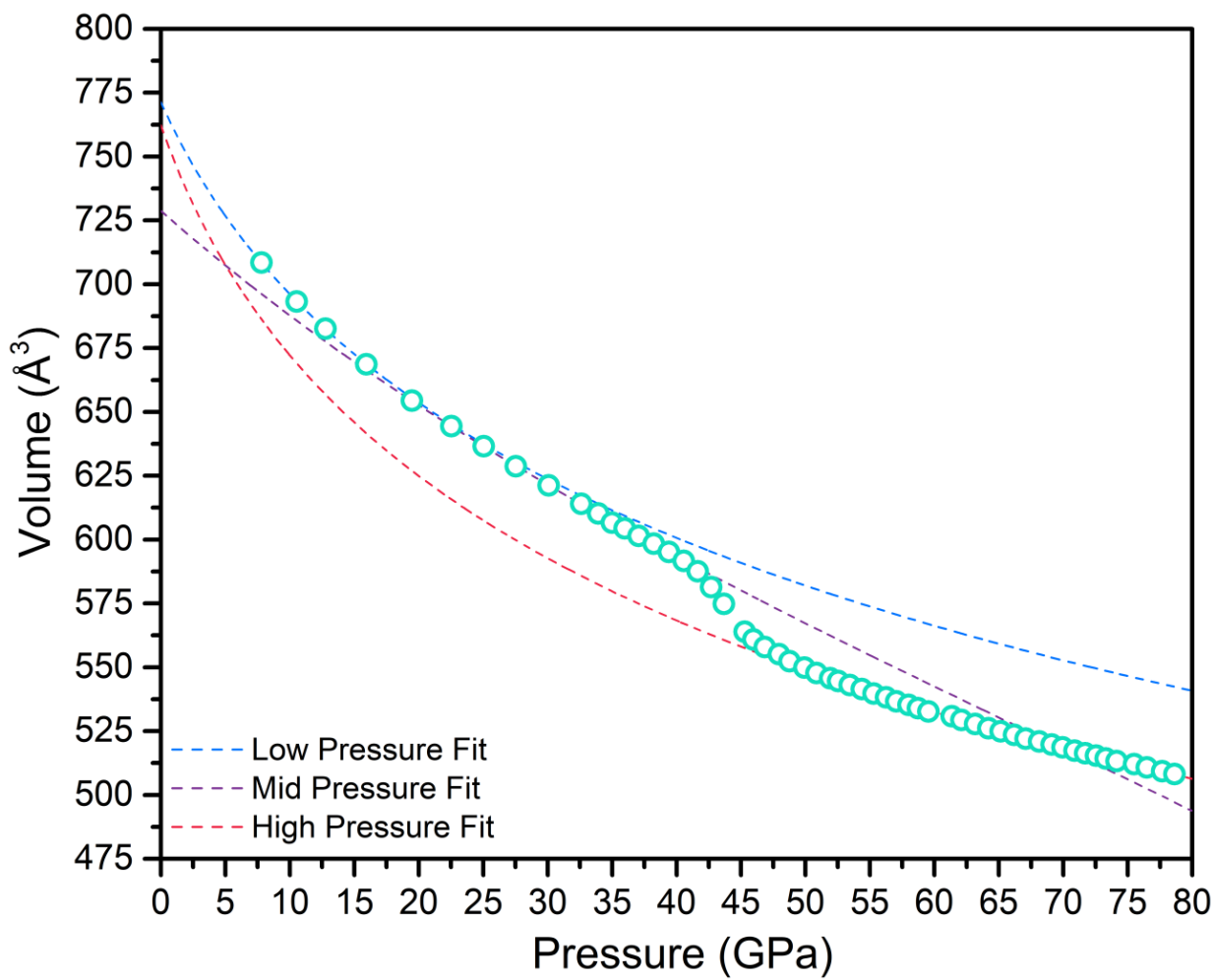

**Figure S4 |** Unit cell volume for jarosite as a function of pressure obtained from fits of the PXRD data. Three equation of state curves fit these data, as denoted by the legend. The low-pressure range fit is from 7.8(1) GPa to 15.9(1) GPa. The mid-pressure range fit covers 19.5(1) to 39.4(3) GPa. The high-pressure range fits 49.9(5) to 78.6(7) GPa. The parameters for the equation of state curves are tabulated in Table S1.

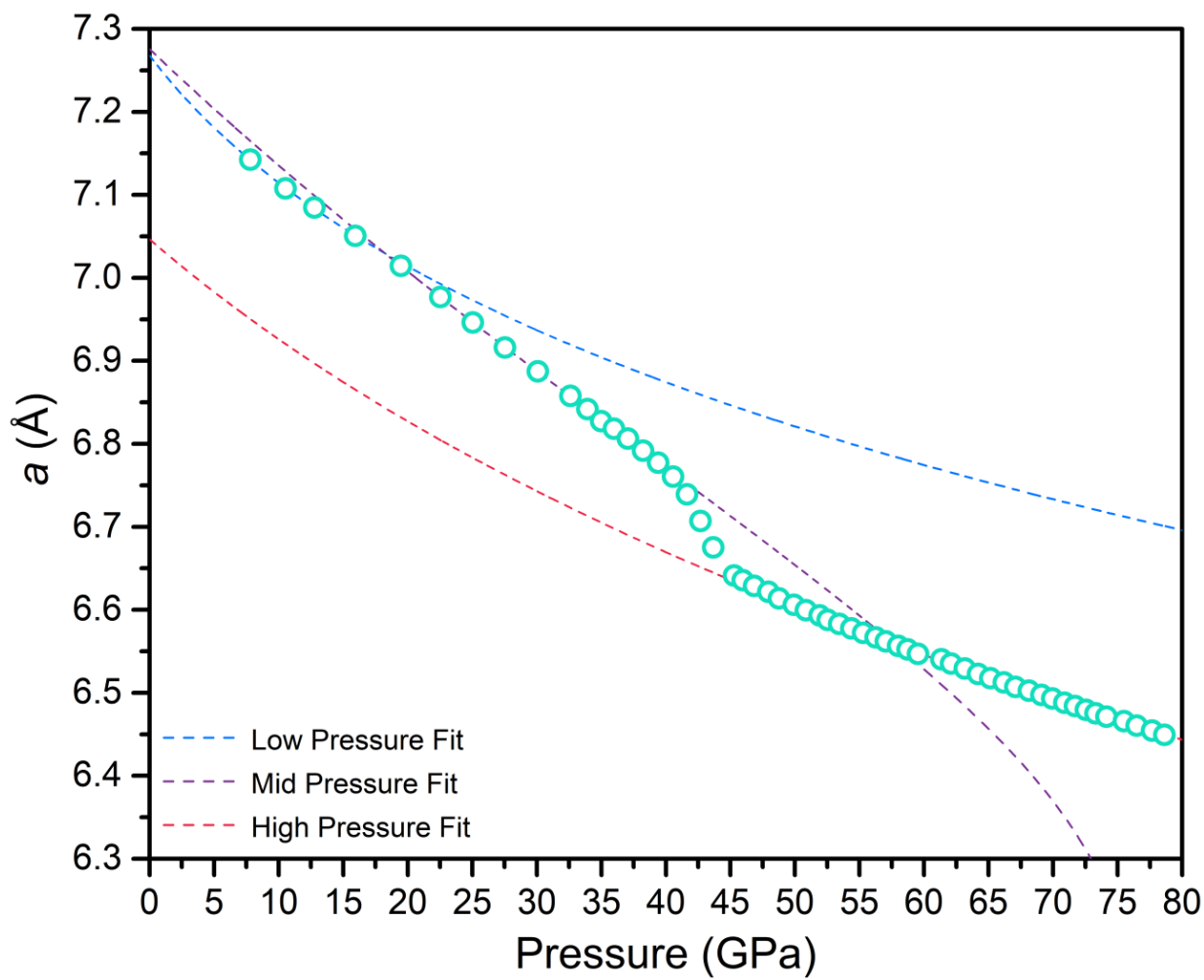

**Figure S5** | *a*-axis length for jarosite as a function of pressure obtained from fits of the PXRD data. Three equation of state curves fit these data, as denoted by the legend. The low-pressure range fit is from 7.8(1) GPa to 15.9(1) GPa. The mid-pressure range fit covers 19.5(1) to 39.4(3) GPa. The high-pressure range fits 49.9(5) to 78.6(7) GPa. The parameters for the equation of state curves are tabulated in Table S1.

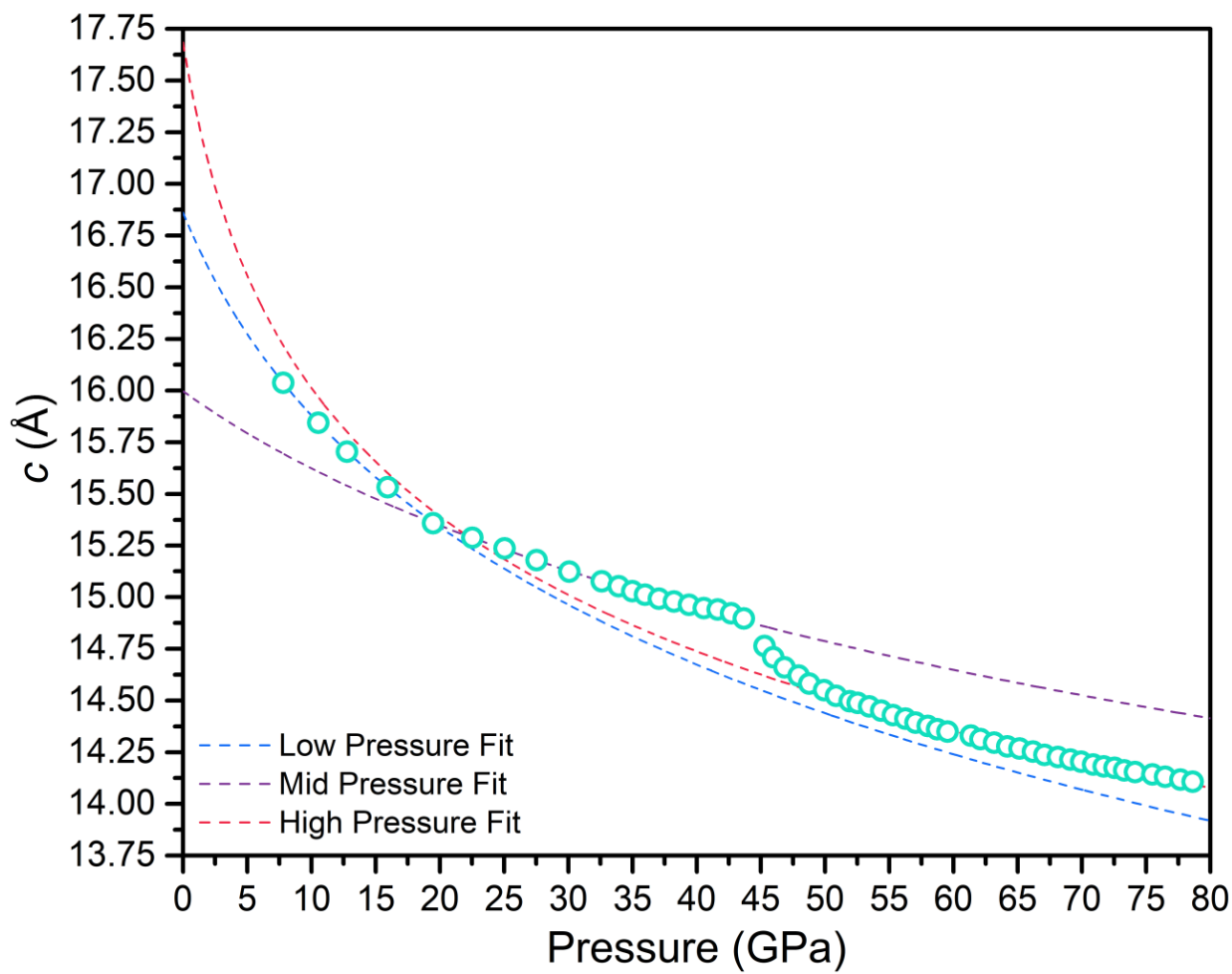

**Figure S6** | *c*-axis length for jarosite as a function of pressure obtained from fits of the PXRD data. Three equation of state curves fit these data, as denoted by the legend. The low-pressure range fit is from 7.8(1) GPa to 15.9(1) GPa. The mid-pressure range fit covers 19.5(1) to 39.4(3) GPa. The high-pressure range fits 49.9(5) to 78.6(7) GPa. The parameters for the equation of state curves are tabulated in Table S1.

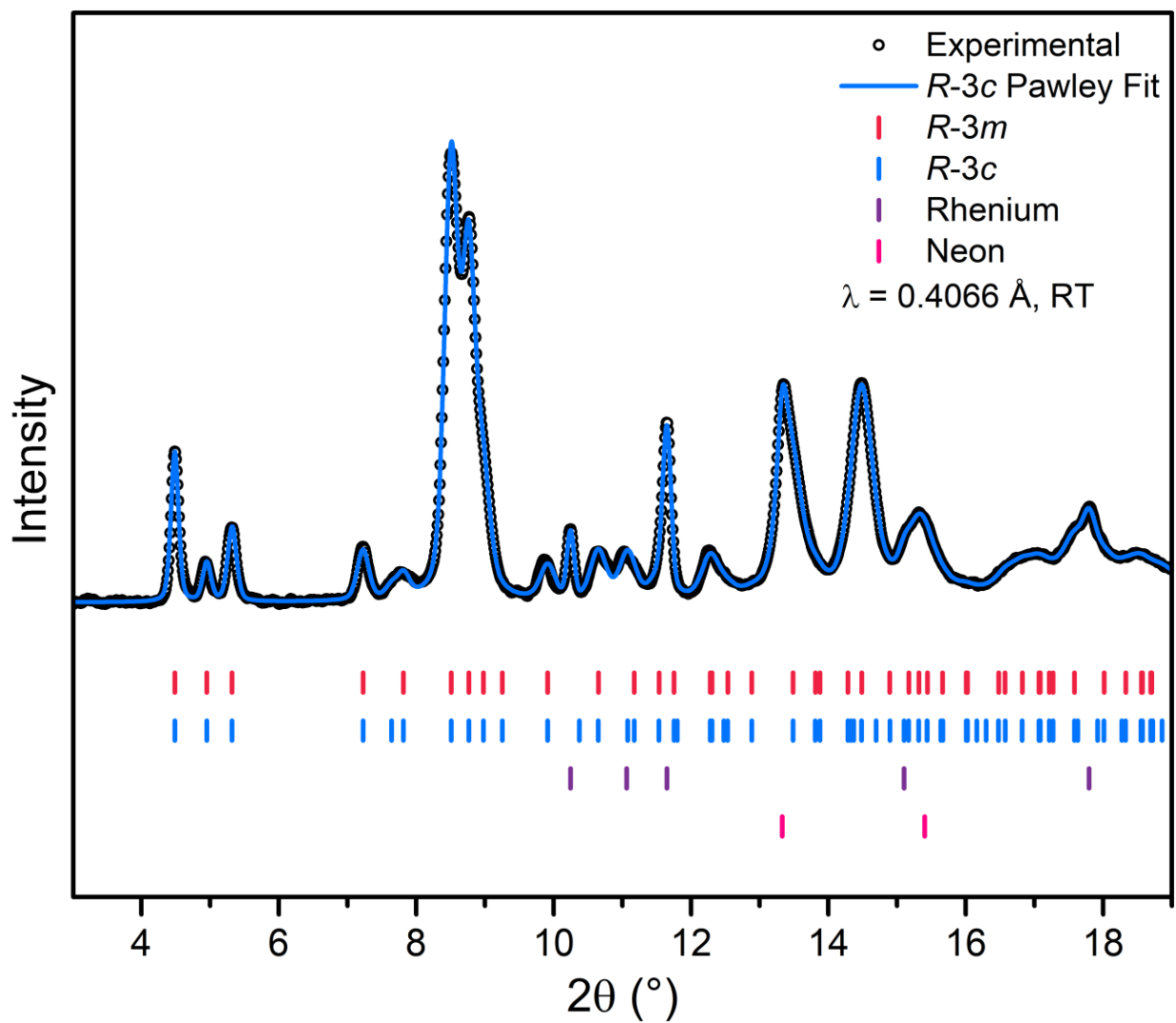

**Figure S7** | Pawley fit of the PXRD pattern obtained at 78.6(7) GPa and ambient temperature. The fit comprises three phases, the calculated $R\bar{3}c$ jarosite structure, rhenium, and neon. The hkl values for these phases are shown below the pattern as well as the hkl values for the $R\bar{3}m$ model. The only new, relatively unobscured hkl in the $R$-3$c$ phase compared to the $R\bar{3}m$ phase occurs at ~7.6 degrees. This region was closely monitored through the phase transition pressure but the difference in the patterns is negligible when accounting for pressure induced peak shifts. At high pressure, the $R\bar{3}c$ model for jarosite yields marginally better fit statistics for the Pawley fits of the PXRD data.

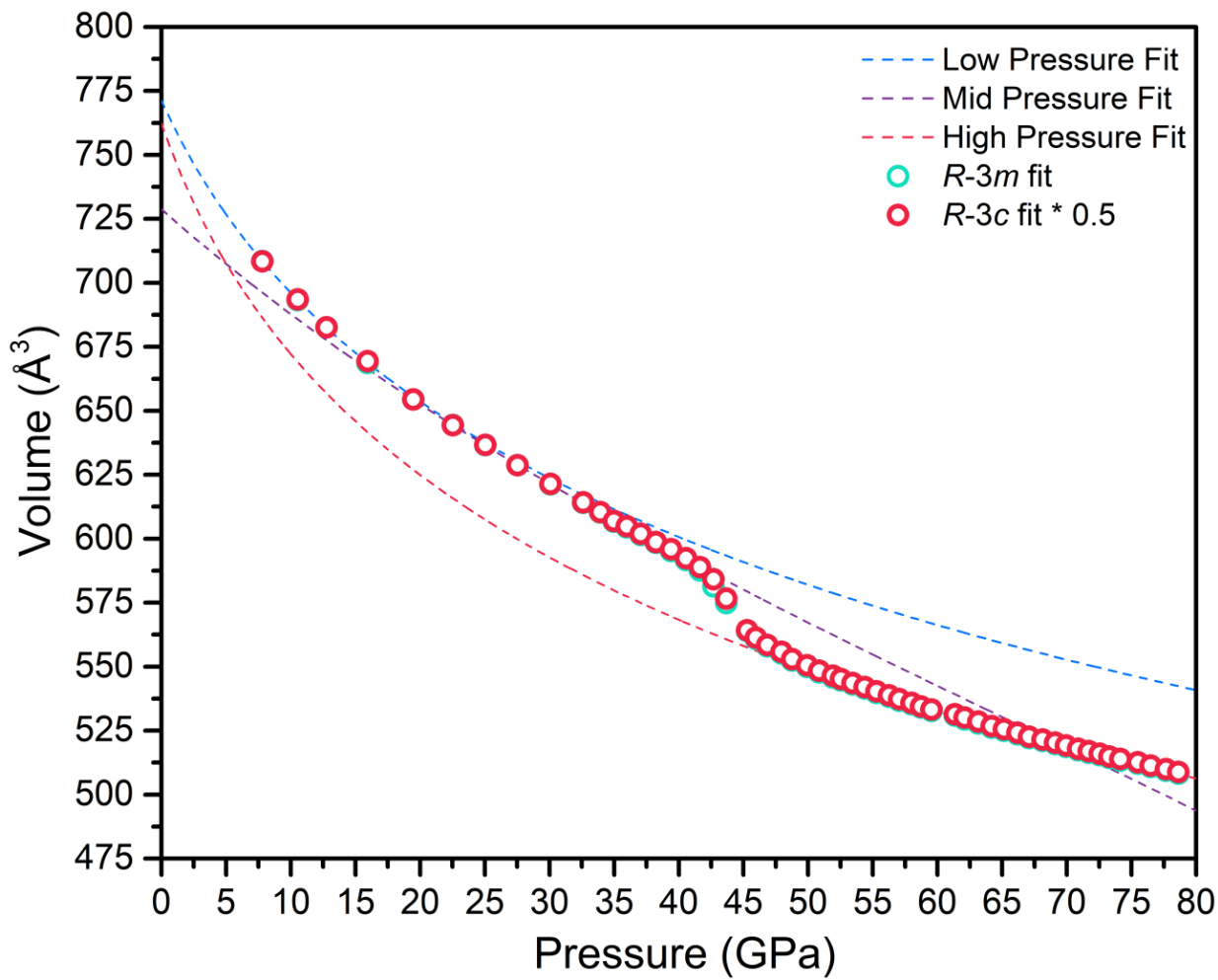

**Figure S8** | The unit cell volume for jarosite as a function of pressure obtained from Pawley fits of the PXRD data using the calculated $R\bar{3}c$ space group is compares well with the values obtained from Pawley fitting the data using the ambient pressure $R\bar{3}m$ space group model. Note that the *c*-axis doubles in the $R\bar{3}c$ phase. As such, the data are multiplied by half for comparison here. As in figure S3, three equation of state curves fit these data, as denoted by the legend. Symbols are commensurate with their error bars, which represent one standard deviation.

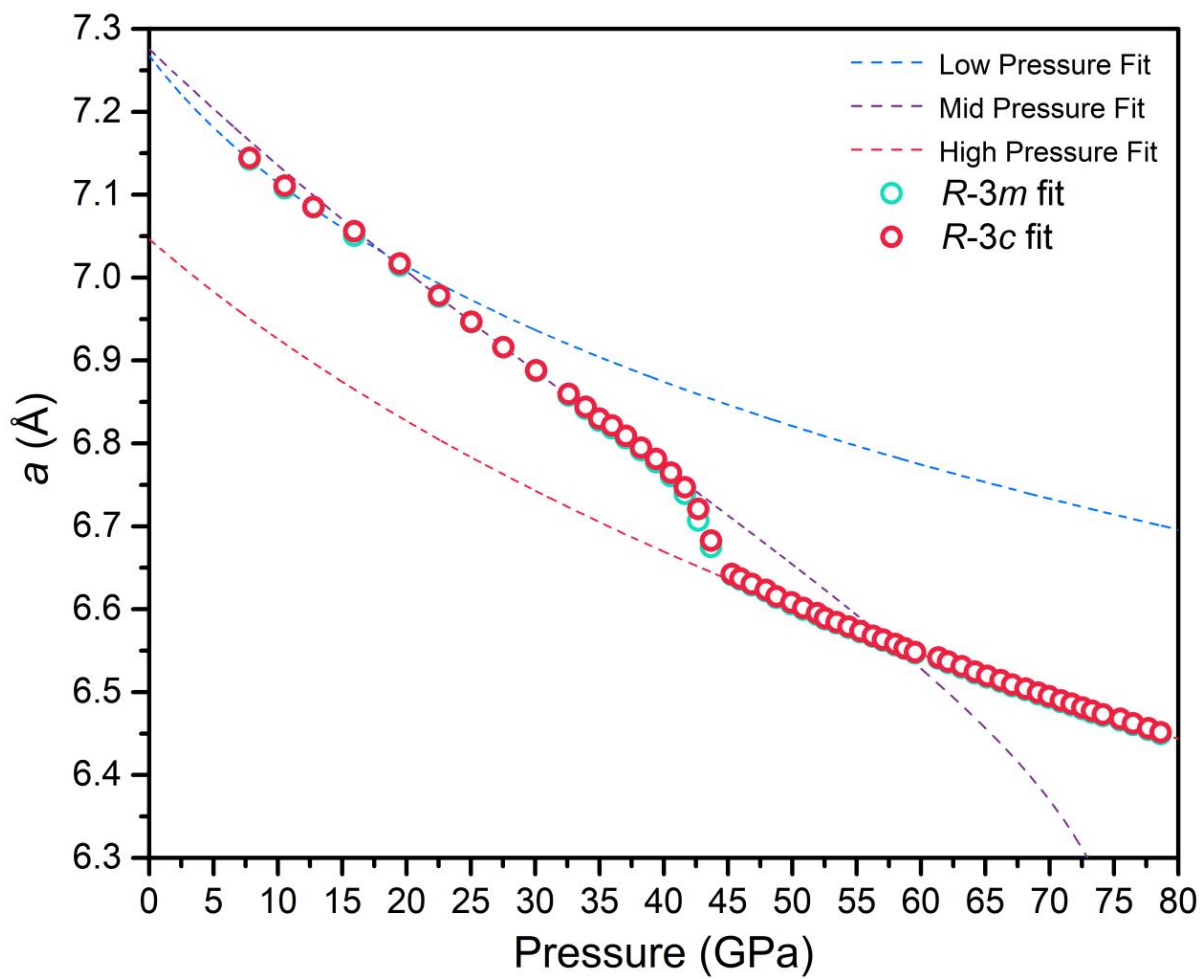

**Figure S9** | The *a*-axis length for jarosite as a function of pressure obtained from Pawley fits of the PXRD data using the calculated $R\bar{3}c$ space group is within error of the values obtained from Pawley fitting the data using the ambient pressure $R\bar{3}m$ space group model. As in figure S4, three equation of state curves fit these data, as denoted by the legend. Symbols are commensurate with their error bars, which represent one standard deviation.

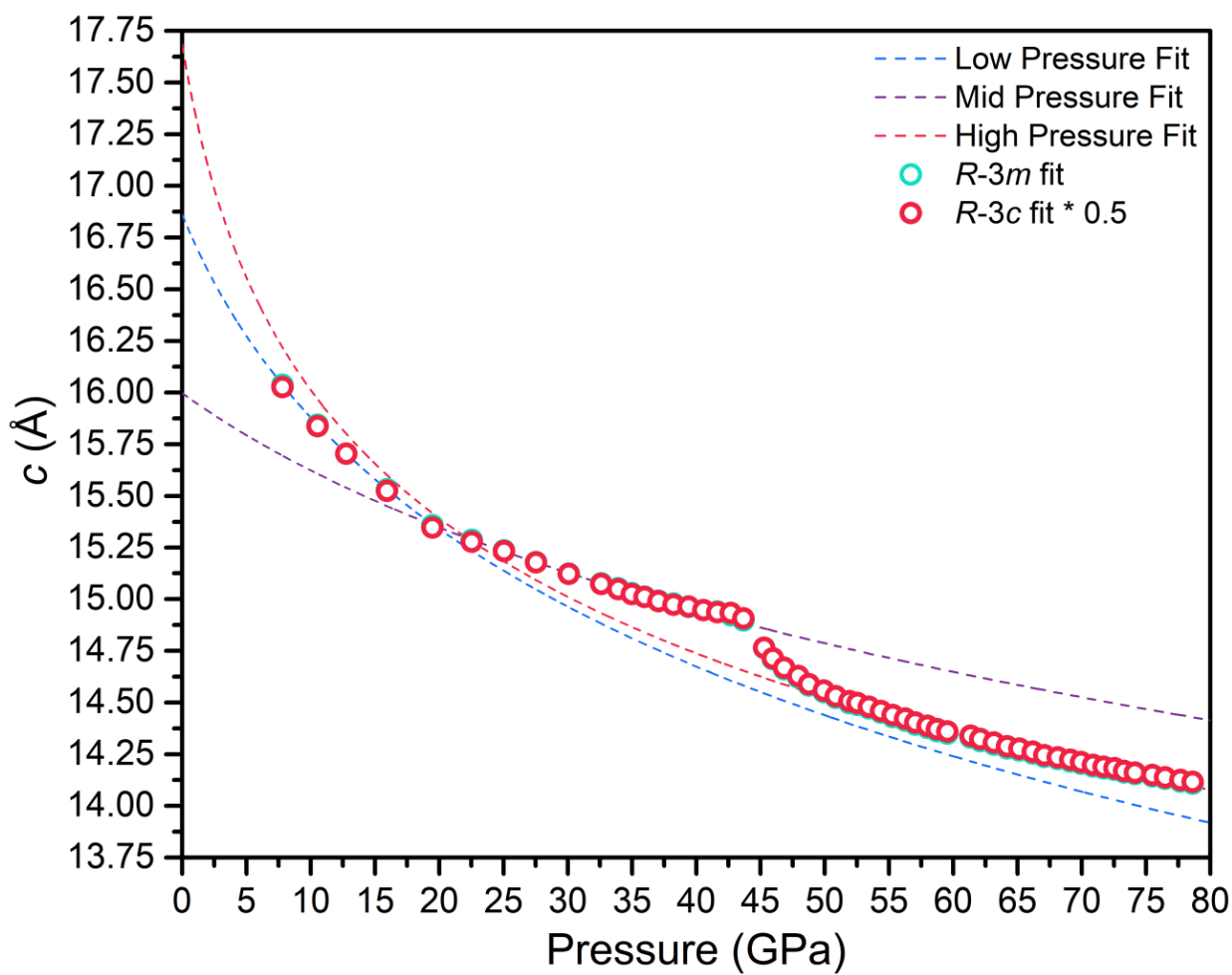

**Figure S10** | The *c*-axis length for jarosite as a function of pressure obtained from Pawley fits of the PXRD data using the calculated $R\bar{3}c$ space group is compares well with the values obtained from Pawley fitting the data using the ambient pressure $R\bar{3}m$ space group model. Note that the *c*-axis doubles in the $R\bar{3}c$ phase. As such, the data are multiplied by half for comparison here. As in figure S5, three equation of state curves fit these data, as denoted by the legend. Symbols are commensurate with their error bars, which represent one standard deviation.

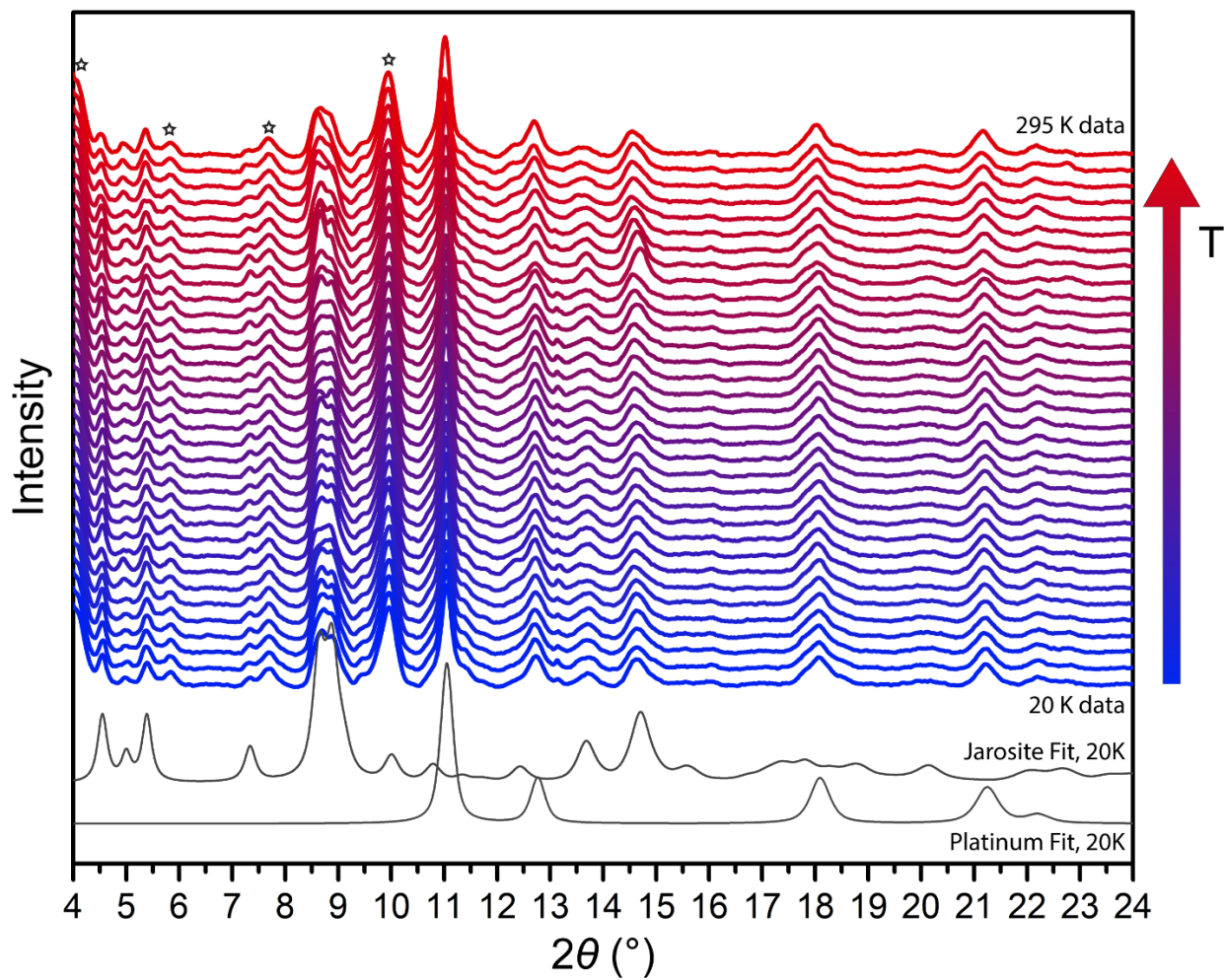

**Figure S11** | Variable temperature PXRD patterns at ~65 GPa were collected at beamline 16-BM-D, HPCAT, APS, ANL. Several phases contribute to the diffraction observed in the patterns, including jarosite, platinum (the pressure calibrant), neon (the pressure-transmitting medium), and the cryostat (diffraction from the cryostat is denoted by stars above the top pattern). Jarosite remains in the same space group and does not undergo any temperature induced phase transitions at ~65 GPa down to ~20 K. Below the patterns, the individual refinements for the jarosite and platinum phases at ~65 GPa and ~20 K are shown as references. The pressure, temperature, and unit cell parameters are given in Table S3.

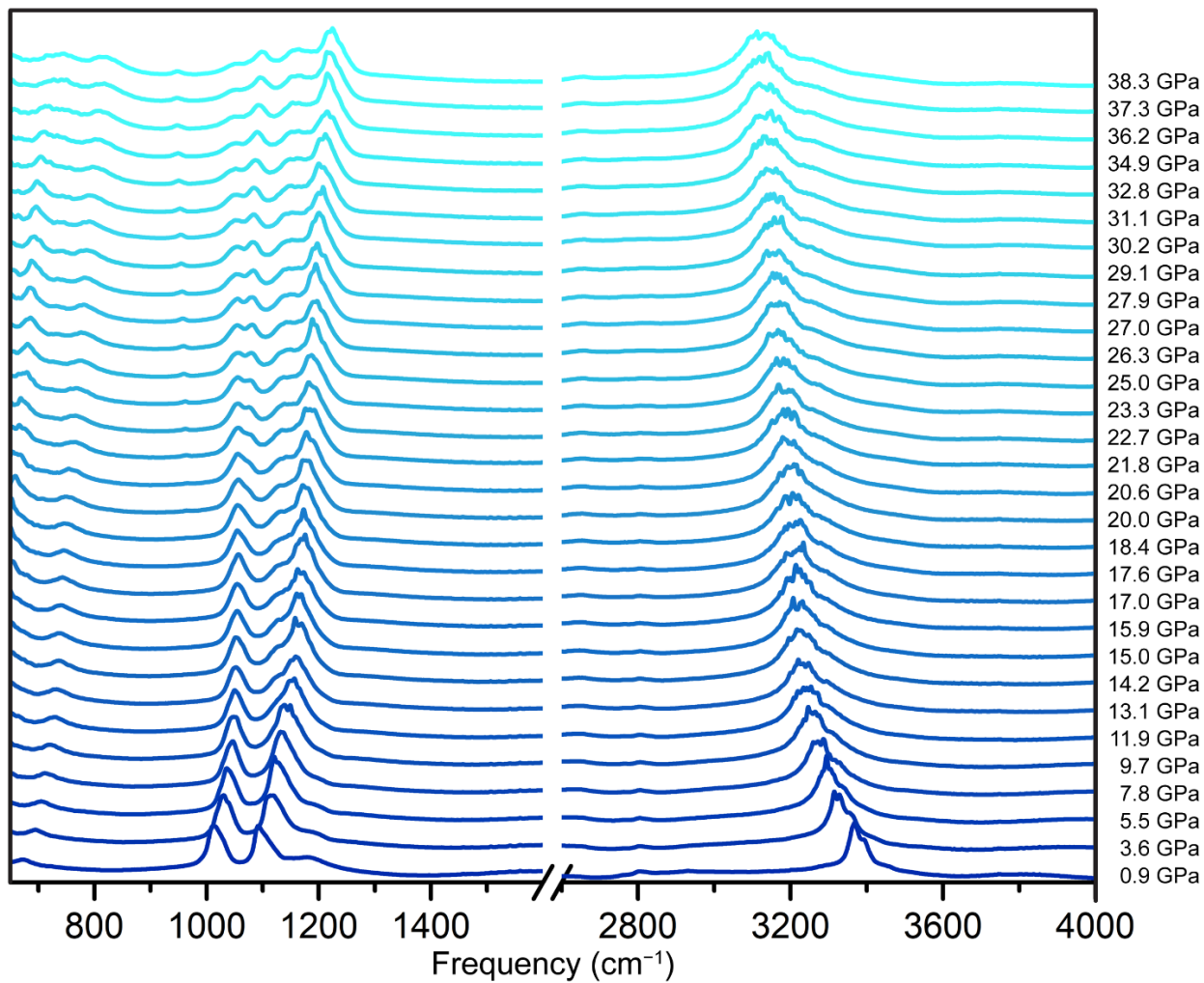

**Figure S12 |** Variable pressure FTIR data collected on a pressed crystalline sample of jarosite in a DAC that employed 300 μm culet diamonds. Saturation of select stretching frequencies is visible. The spectra are not modified. For this sample, neon acted as the pressure transmitting medium, and the $R_1$ fluorescence feature of ruby acted as the pressure calibrant.

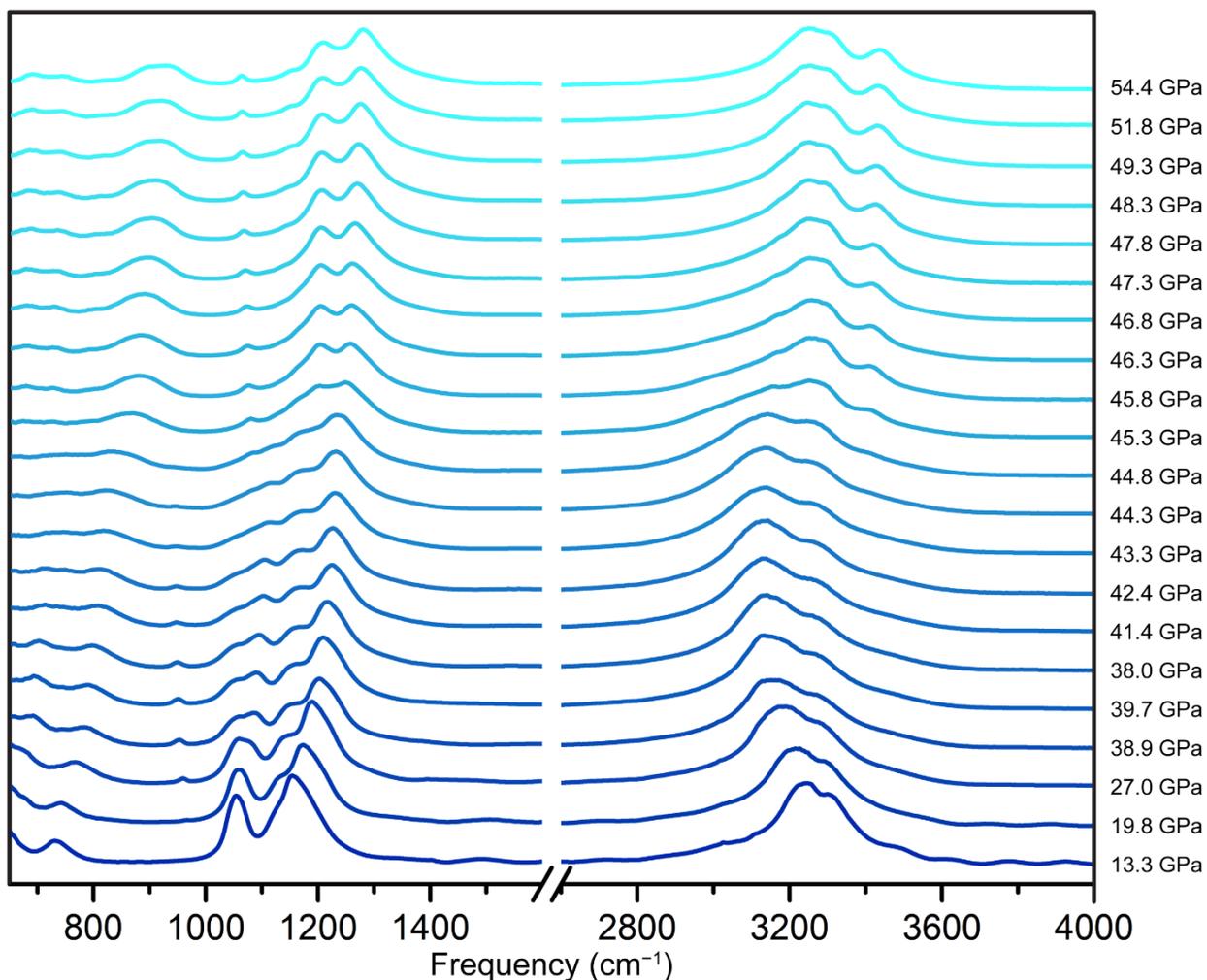

**Figure S13** | Variable pressure FTIR data collected on a pressed crystalline sample of jarosite in a DAC that employed 200 μm culet diamonds. The spectra are not modified. For this sample, KBr acted as the pressure transmitting medium, and the $R_1$ fluorescence feature of ruby acted as the pressure calibrant up to ~35 GPa, at which point the ruby bridged the diamonds. Then, the first order Raman band of the diamonds acted as the calibrant.

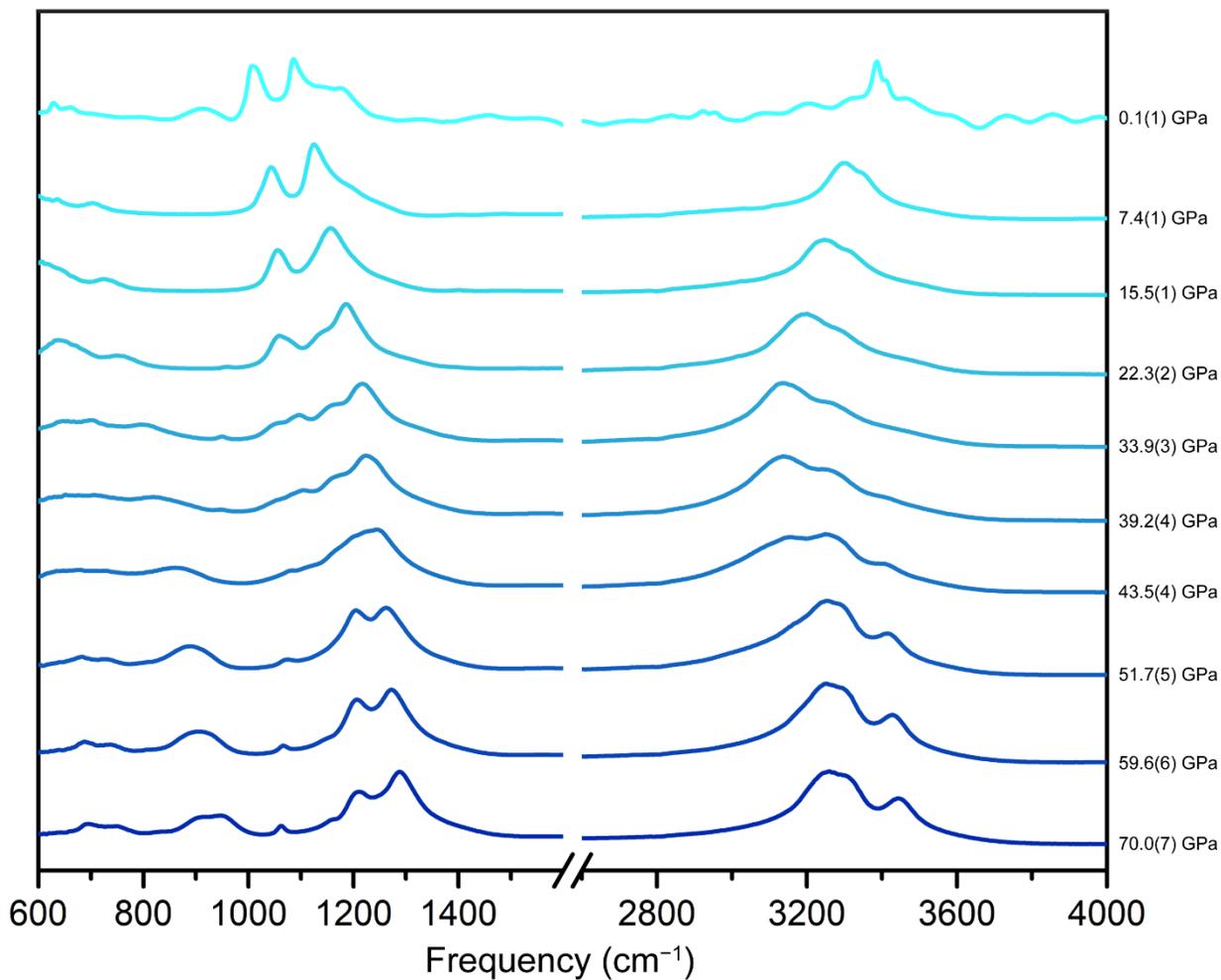

**Figure S14** | Variable pressure FTIR data measured upon decompression of a pressed crystalline sample of jarosite in a DAC that employed 200 μm culet diamonds. The spectra are not modified. For this sample, KBr acted as the pressure transmitting medium the first order Raman band of the diamonds acted as the calibrant. The pressure is noted to the right of each spectrum.

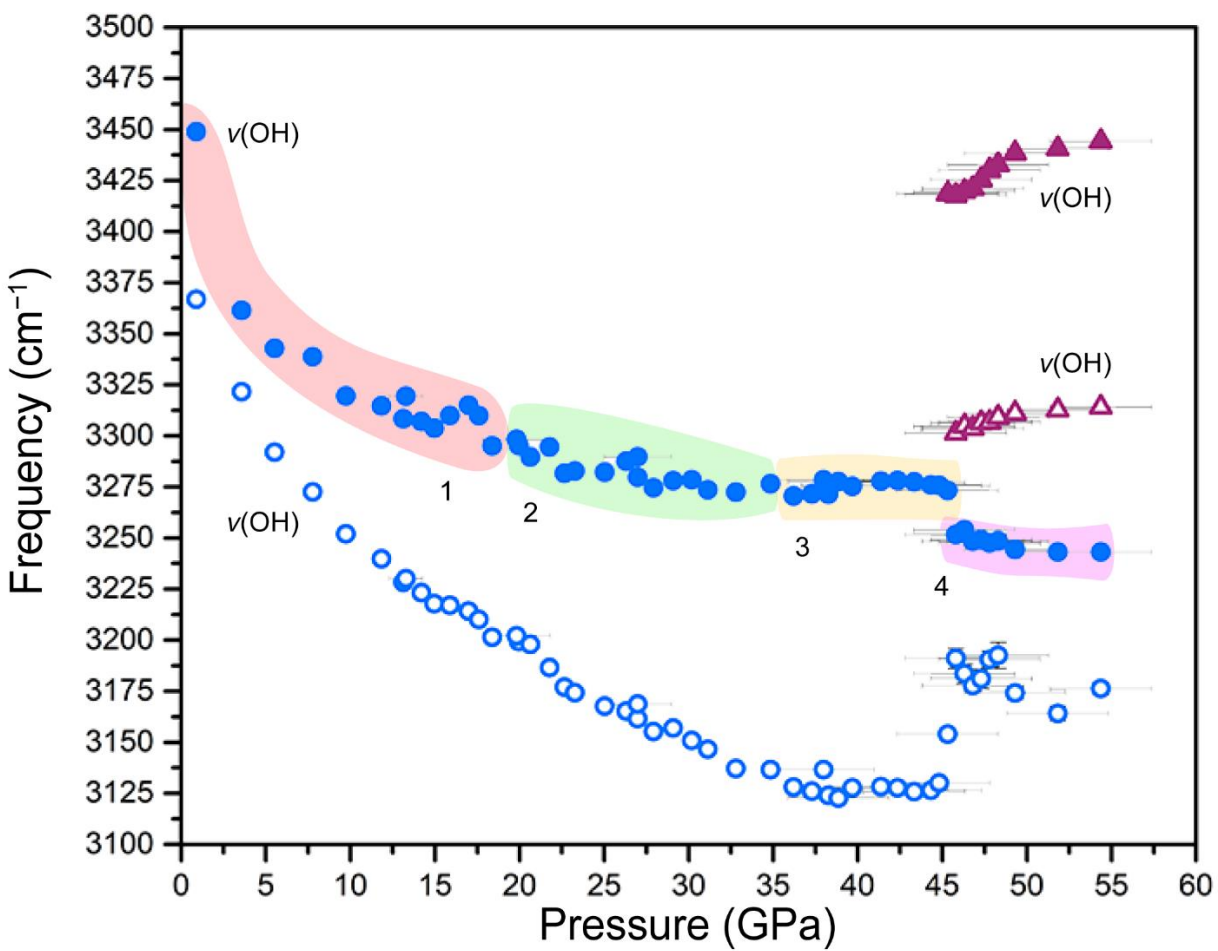

**Figure S15** | The measured frequency of the observable IR active modes in the 3100–3500 cm$^{-1}$ region is plotted as a function of pressure. Error bars are smaller than their symbols in both directions unless shown. Color and symbol shape denote the identity of the distinct modes through the first-order transition, as identified using an analysis of the mode FWHM values as seen in Fig. S18.

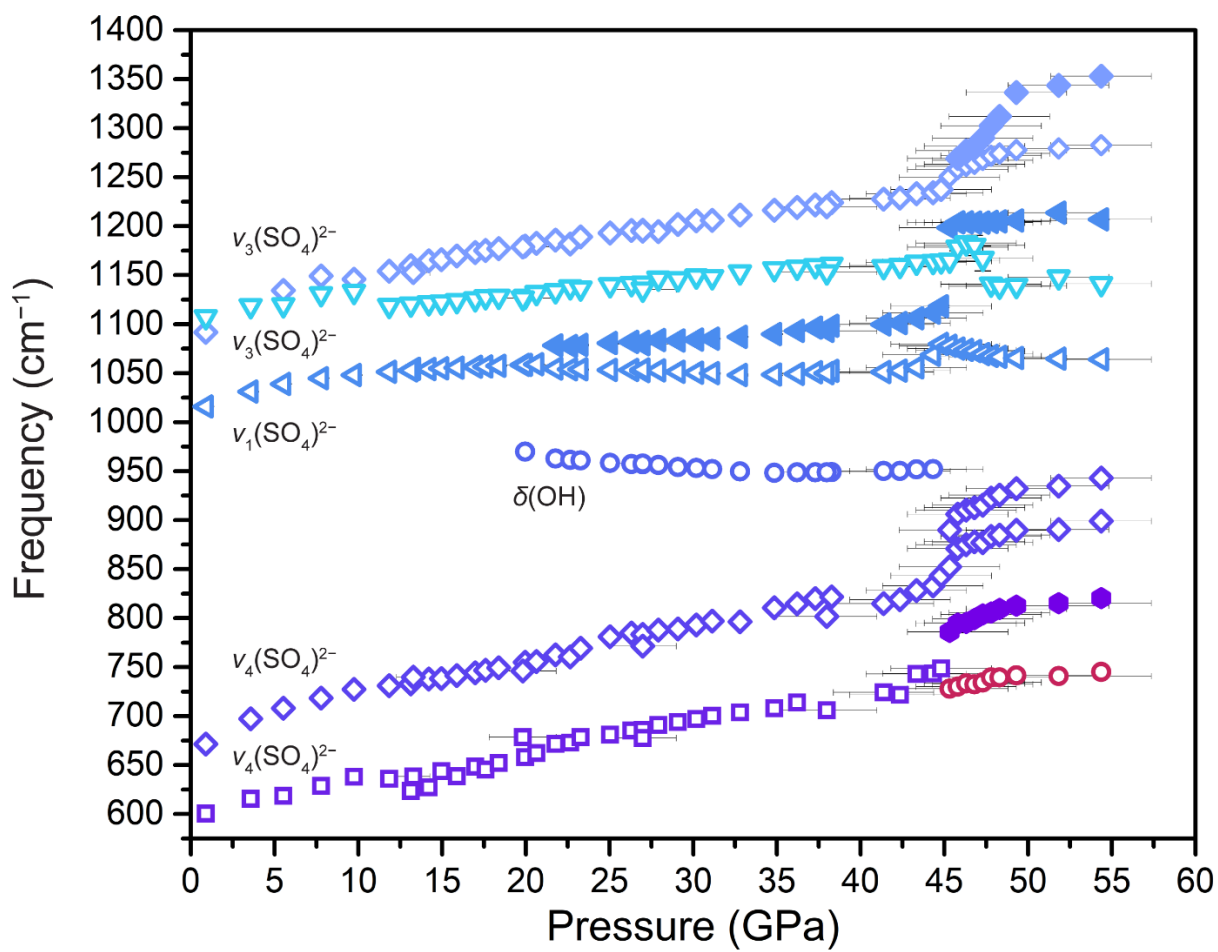

**Figure S16** | The measured frequency of the observable IR active modes in the 600–1400 cm$^{-1}$ region is plotted as a function of pressure. Error bars are smaller than their symbols in both directions unless shown. Color and symbol shape denote the identity of the distinct modes through the first-order transition, as identified using an analysis of the mode FWHM values as seen in Fig. S19.

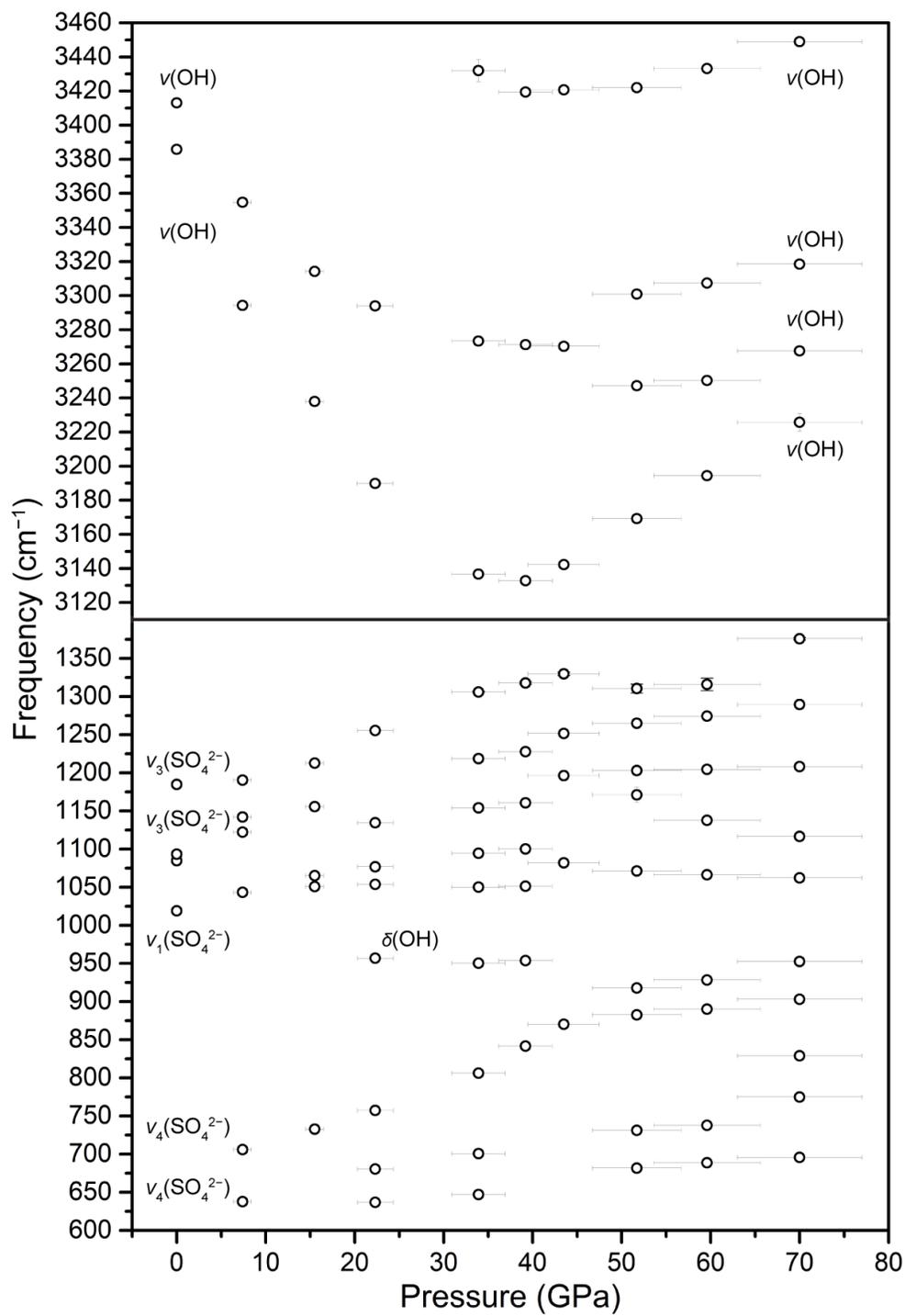

**Figure S17 |** FTIR mode position upon decompression.

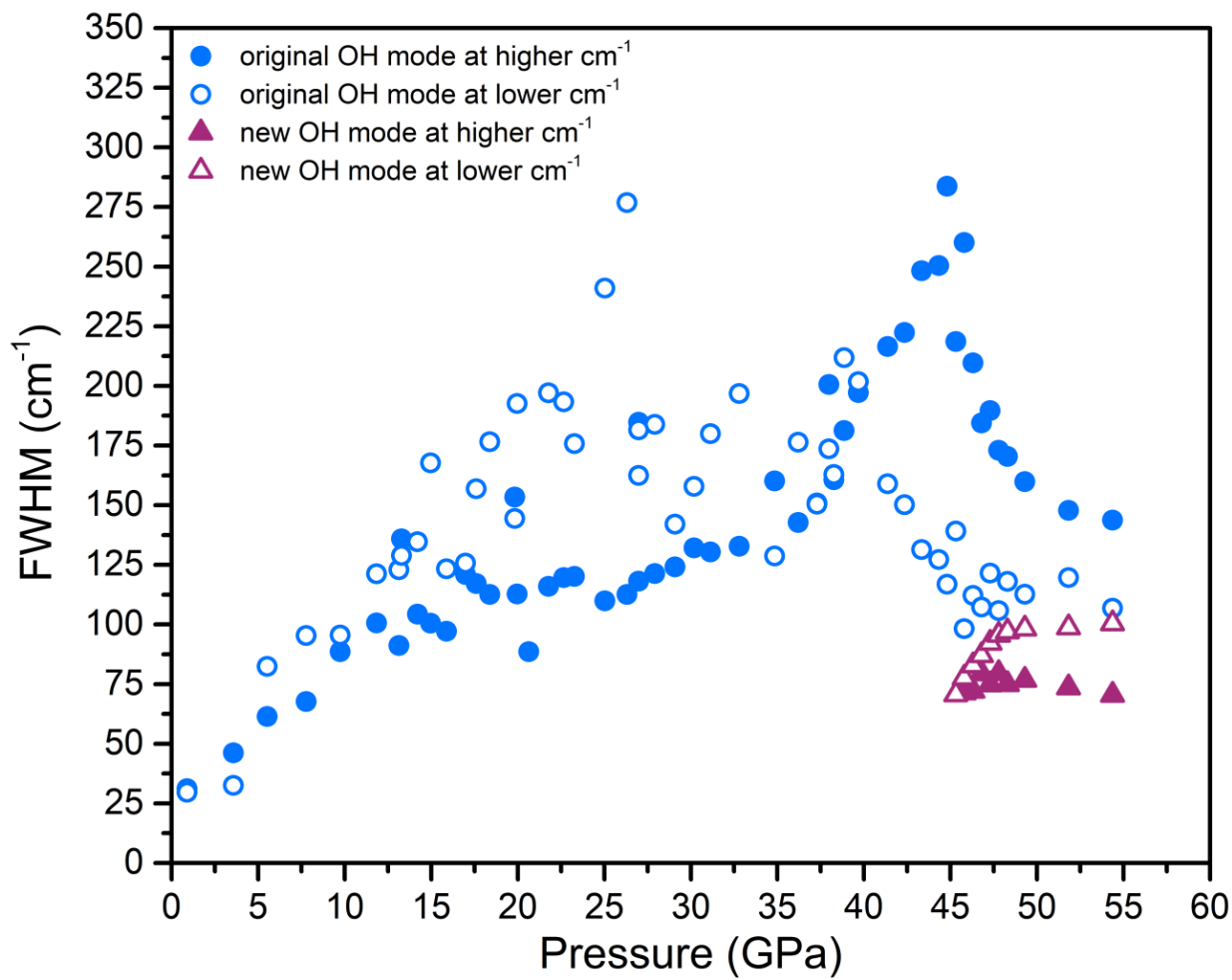

**Figure S18** | The full width at half max (FWHM) values for the $v$(OH) modes are plotted as a function of pressure. The shape, fill, and color of the symbols denoting the FWHM for a specific mode matches with the plot of the frequency of each mode as a function of pressure shown in the main text (Fig. 2). The FWHM values for the two low-pressure $v$(OH) modes are continuous across the phase transition, while the two high-pressure modes appear with FWHM that is distinct and lower in cm$^{-1}$ than the original modes. This analysis enabled us to track the modes through the first order phase transition at 45 GPa.

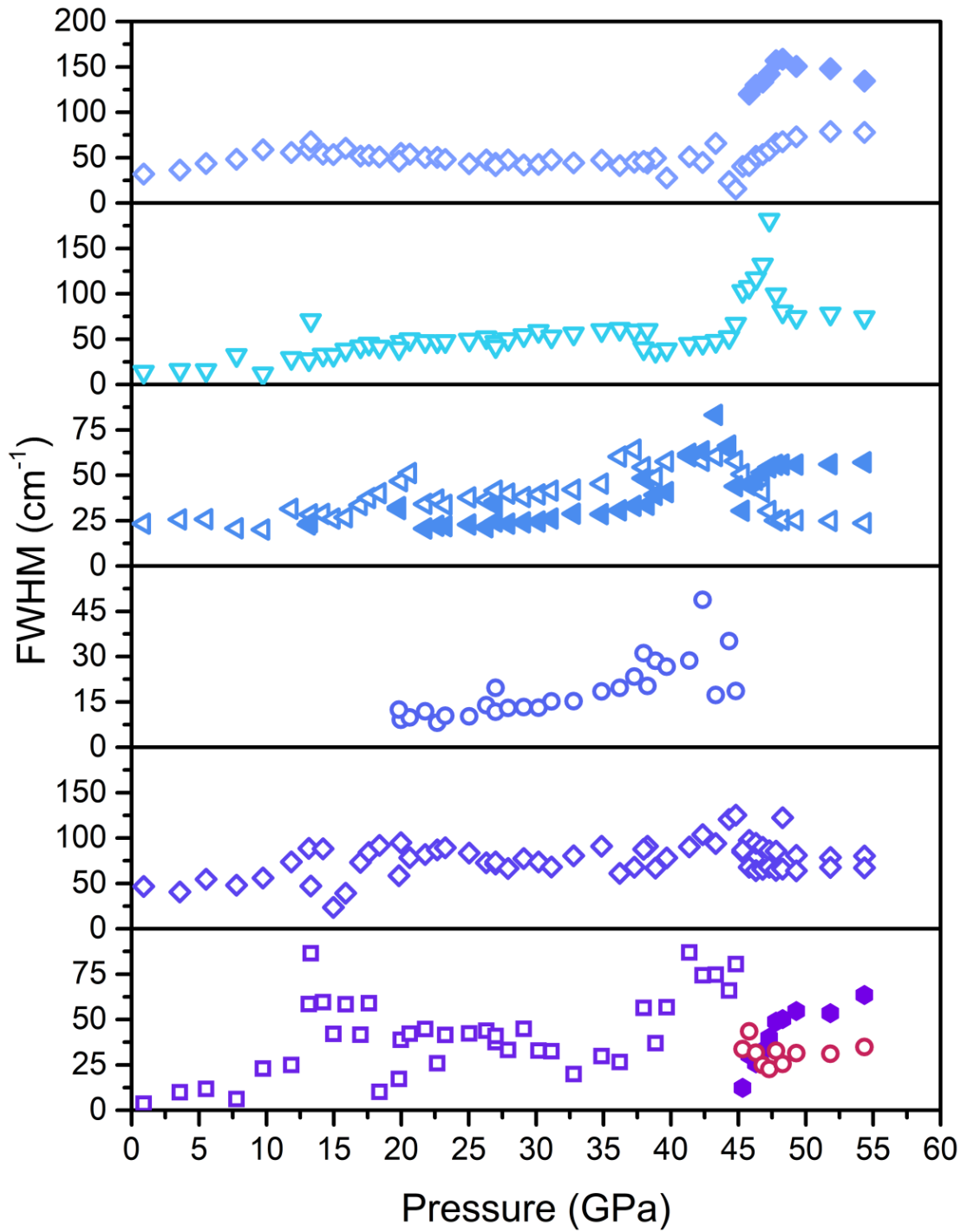

**Figure S19** | The FWHM values for the low-wavenumber modes are plotted as a function of pressure. The shape, fill, and color of the symbols denoting the FWHM for a specific mode matches with the plot of the frequency of each mode as a function of pressure shown in the main text (Fig. 2). The FWHM analysis for this spectral region is more obscured because the sulfate modes have very similar FWHM values and they overlap. To facilitate the analysis, the data are stacked in offset plots instead of presented in one plot.

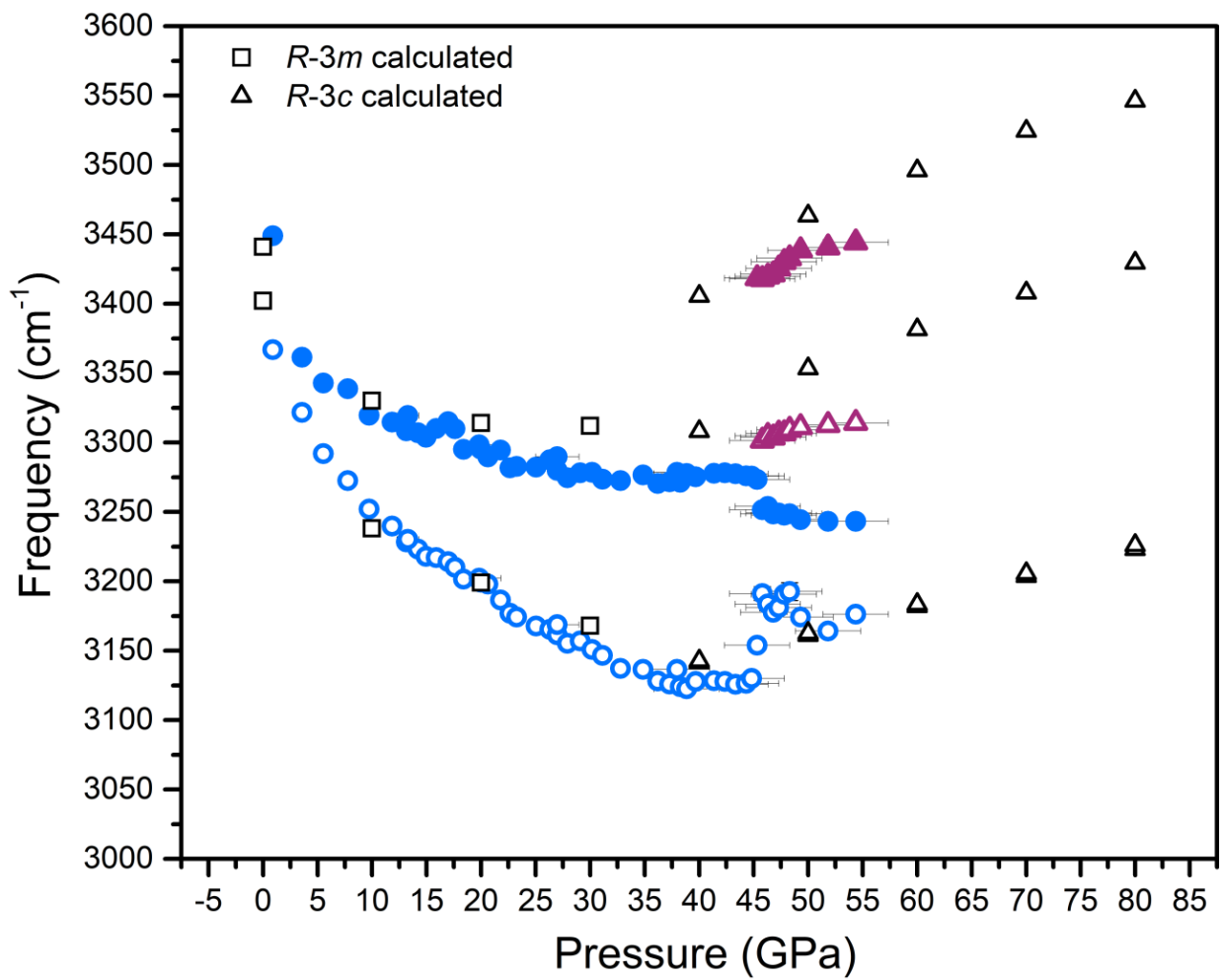

**Figure S20 |** The number and position in frequency as a function of pressure of the calculated FTIR active modes in the high-wavenumber region agree with the experimentally observed modes. Note that there are four calculated modes above the phase transition. Two modes nearly overlap between ~3100 and ~3250 wavenumbers.

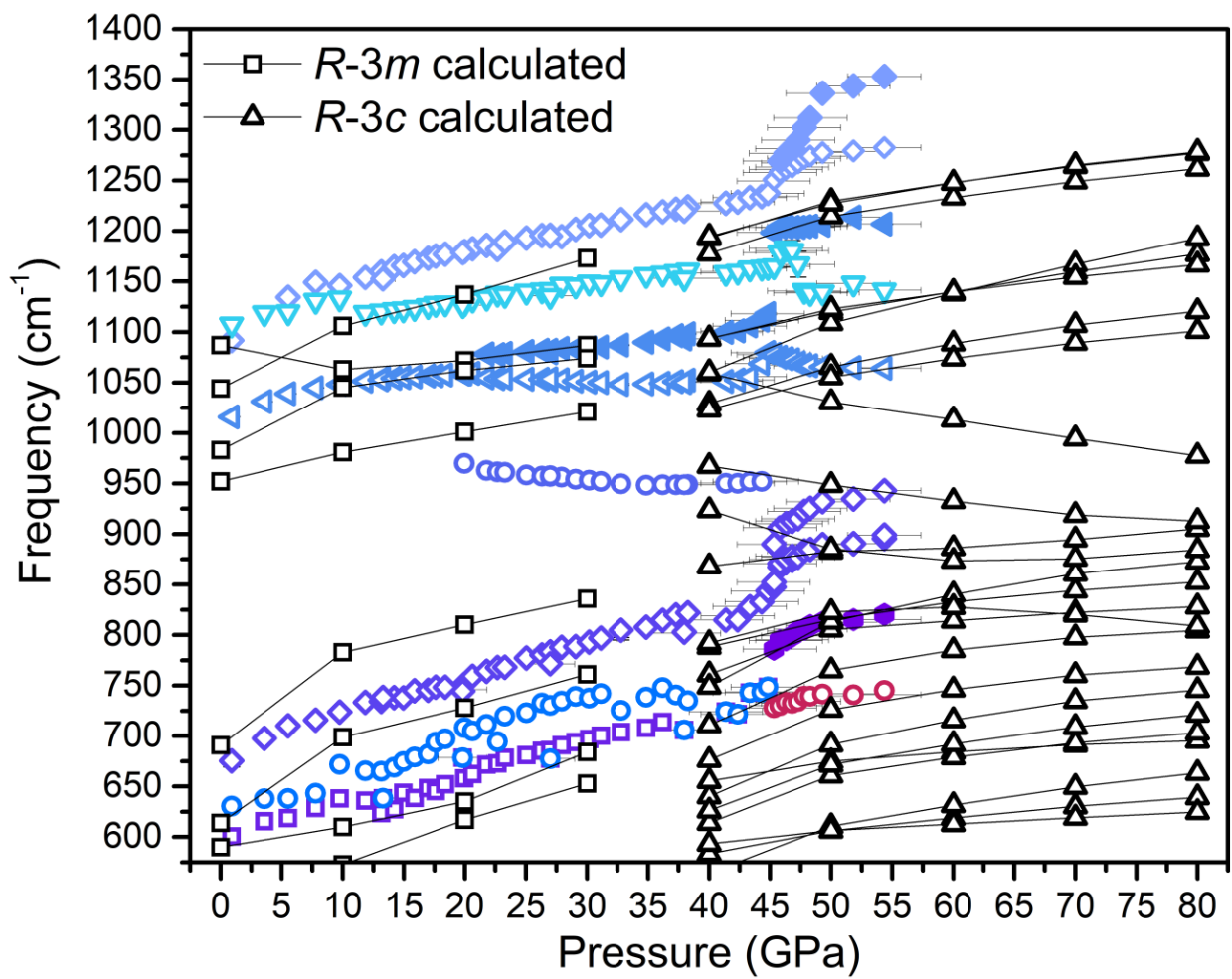

**Figure S21** | The number and position in frequency as a function of pressure of the calculated FTIR active modes in the low-wavenumber region agree with the experimentally observed modes.

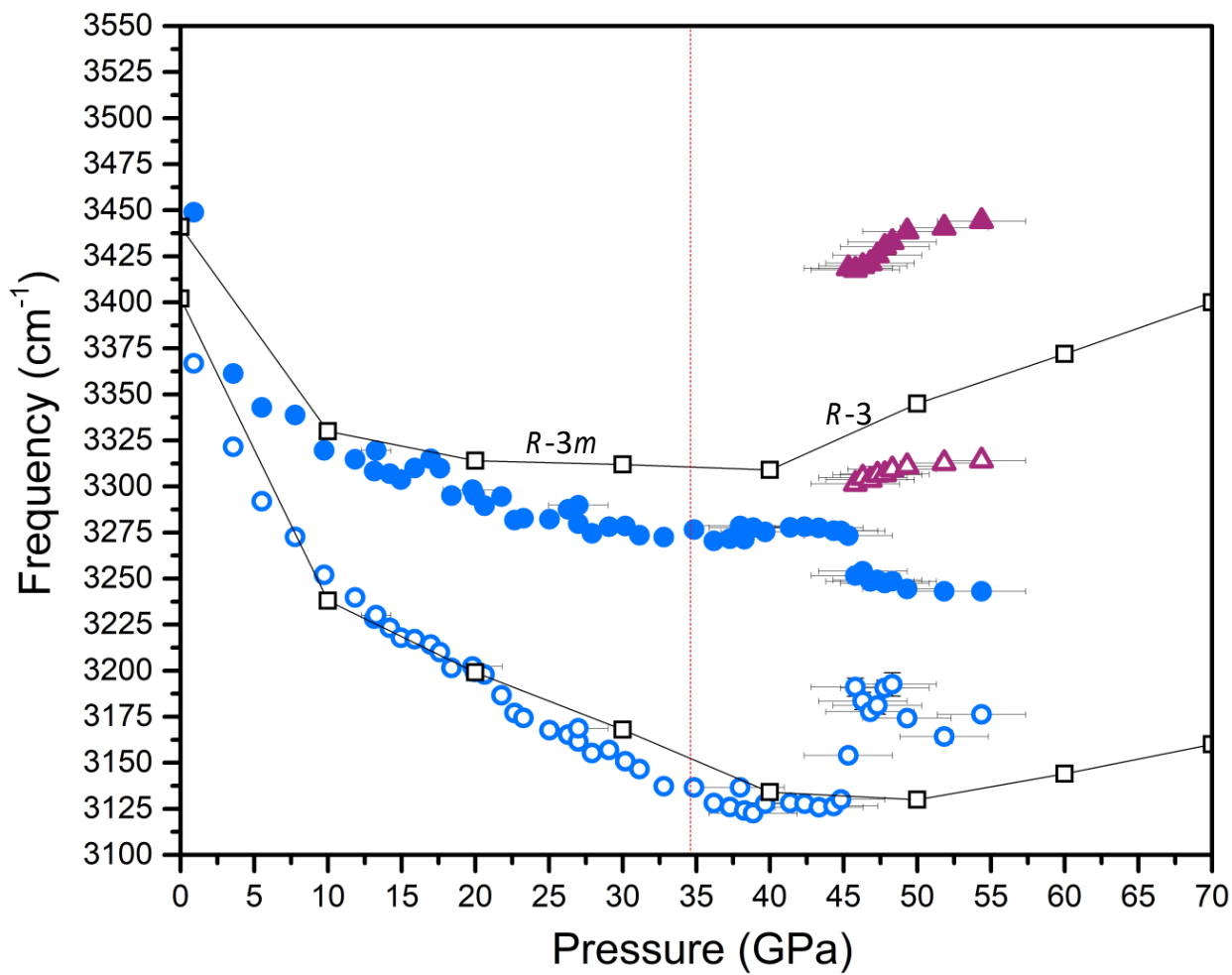

**Figure S22** | The number and position in frequency as a function of pressure of the calculated FTIR active modes for the $R\bar{3}m$ and $R\bar{3}$ structures in the high-wavenumber region do not agree with the experimentally observed modes.

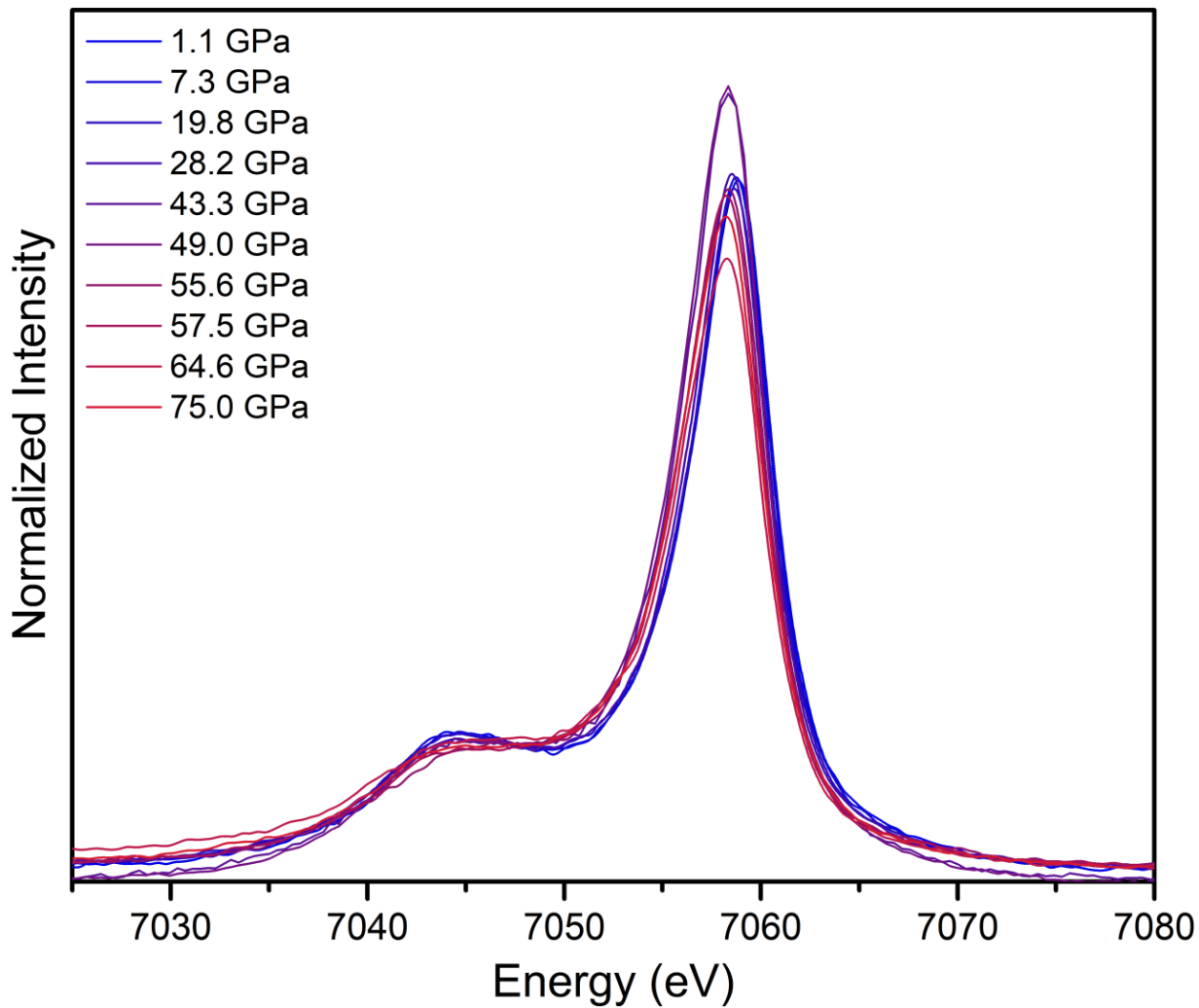

**Figure S23** | x-ray emission spectroscopy spectra at various pressures for jarosite. This figure highlights the similarity between the low- and high-pressure data. As pressure increases, the Kβ′ feature decreases in intensity slightly, the valley between the two peaks increases in intensity, and the K$\beta_{1,3}$ peak shifts to lower energy and decreases in intensity slightly. The plot of the IAD values in Figure S7, which includes previously published data for this material, reflects the change in jarosite's electronic structure as a function of pressure up to 75.0(7) GPa. This change is caused by a linear increase in the Fe–O bond covalency. These data were collected from two different samples of jarosite in two separate experiments. The first experiment included all of the data except for the spectra at 43.3 and 49.0 GPa, which came from the second run. Differences in single crystal orientation or in background noise (and therefore the integral normalization) across the experiments may account for slight differences in intensity at the K$\beta_{1,3}$ peak. The IAD analysis (Figure S8) allows for a quantitative, direct comparison of the spectra.

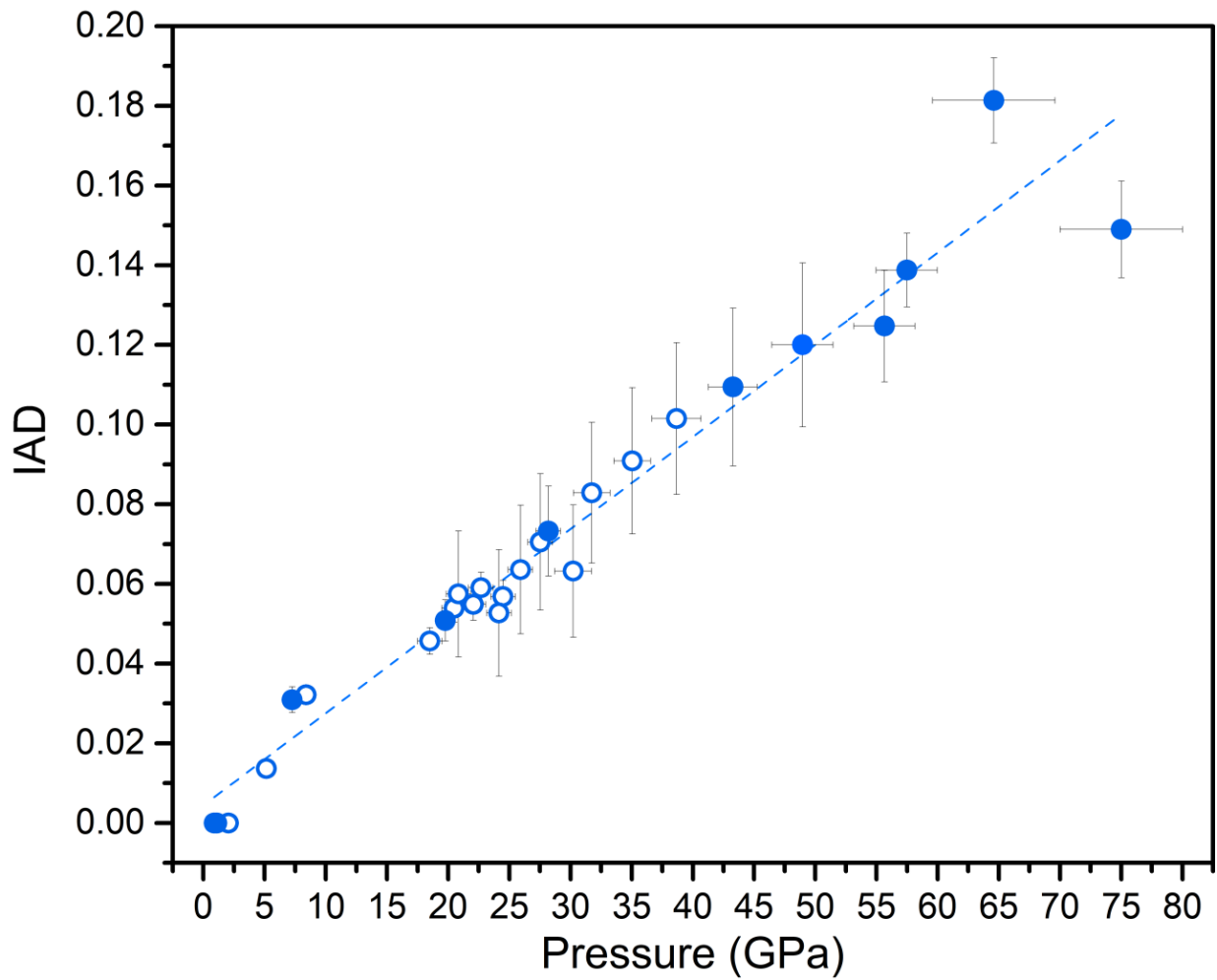

**Figure S24** | The Integral of the Absolute value of the Difference curve (IAD) values for the XES spectra collected for jarosite up to 75.0(7) GPa. Data points deriving from spectra not plotted in Figure S6 derive from a previously published IAD analysis of jarosite. These data points are hollow, while the filled data points are from the spectra plotted in Figure S6. The dotted line is a linear fit to the aggregate of the data. The slope is 0.0023 IAD units per GPa and $R^2 = 0.956$. The linear relationship illustrates the lack of a phase transition in the electronic structure of $Fe^{3+}$ ions in jarosite up to 75.0(7) GPa.

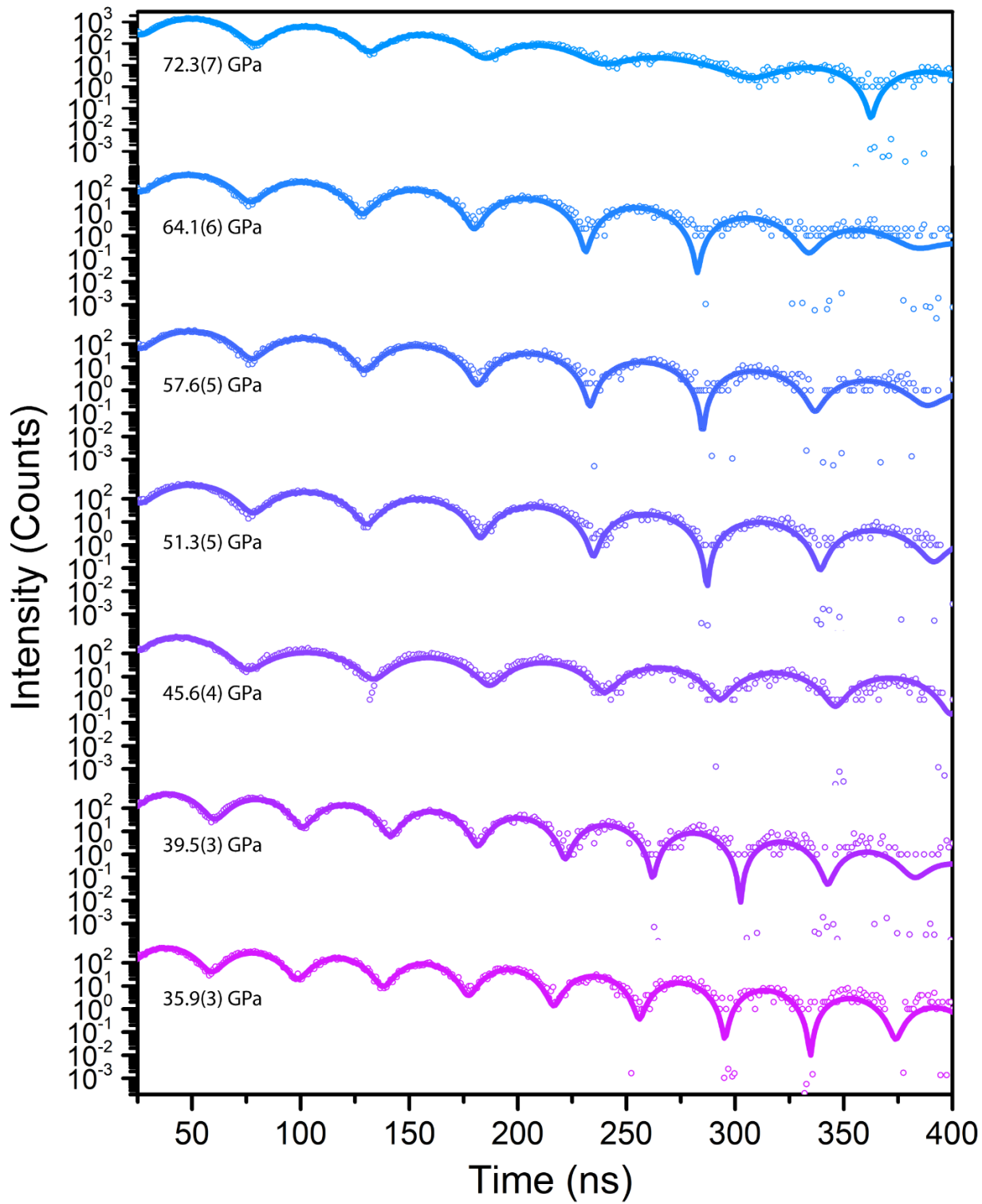

**Figure S25** | The SMS data and fits for jarosite at ambient temperature. The data were collected at Sector 3, APS, ANL, when the synchrotron operated in hybrid bunch mode. The pressure is specified underneath each spectrum. These spectra do not exhibit magnetic hyperfine splitting. The pressures and fit parameters are listed in Table S5. There is a discontinuity in spectra shape between 39.5(3) GPa and 45.6(4) GPa corresponding to the decrease in $\Delta E_Q$.

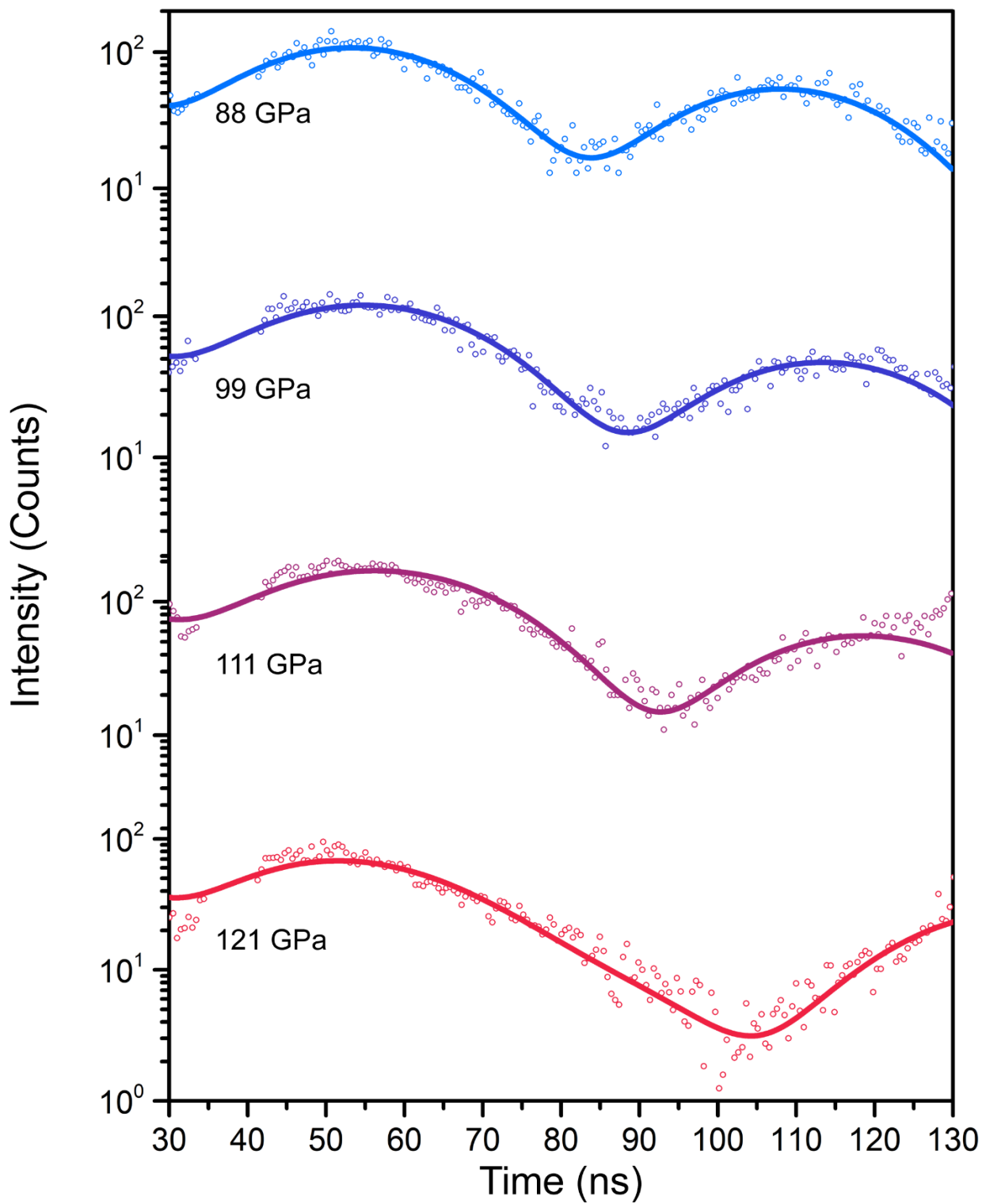

**Figure S26** | The SMS data and fits for jarosite at ambient temperature collected up to a megabar. The data were collected at beamline 16-ID-D, HPCAT, APS, ANL. The pressure is specified underneath each spectrum. Spurious bunches led to non-physical features in the spectra that are masked between 35 and 41 ns. These spectra do not exhibit magnetic hyperfine splitting. The pressures and fit parameters are listed in Table S5.

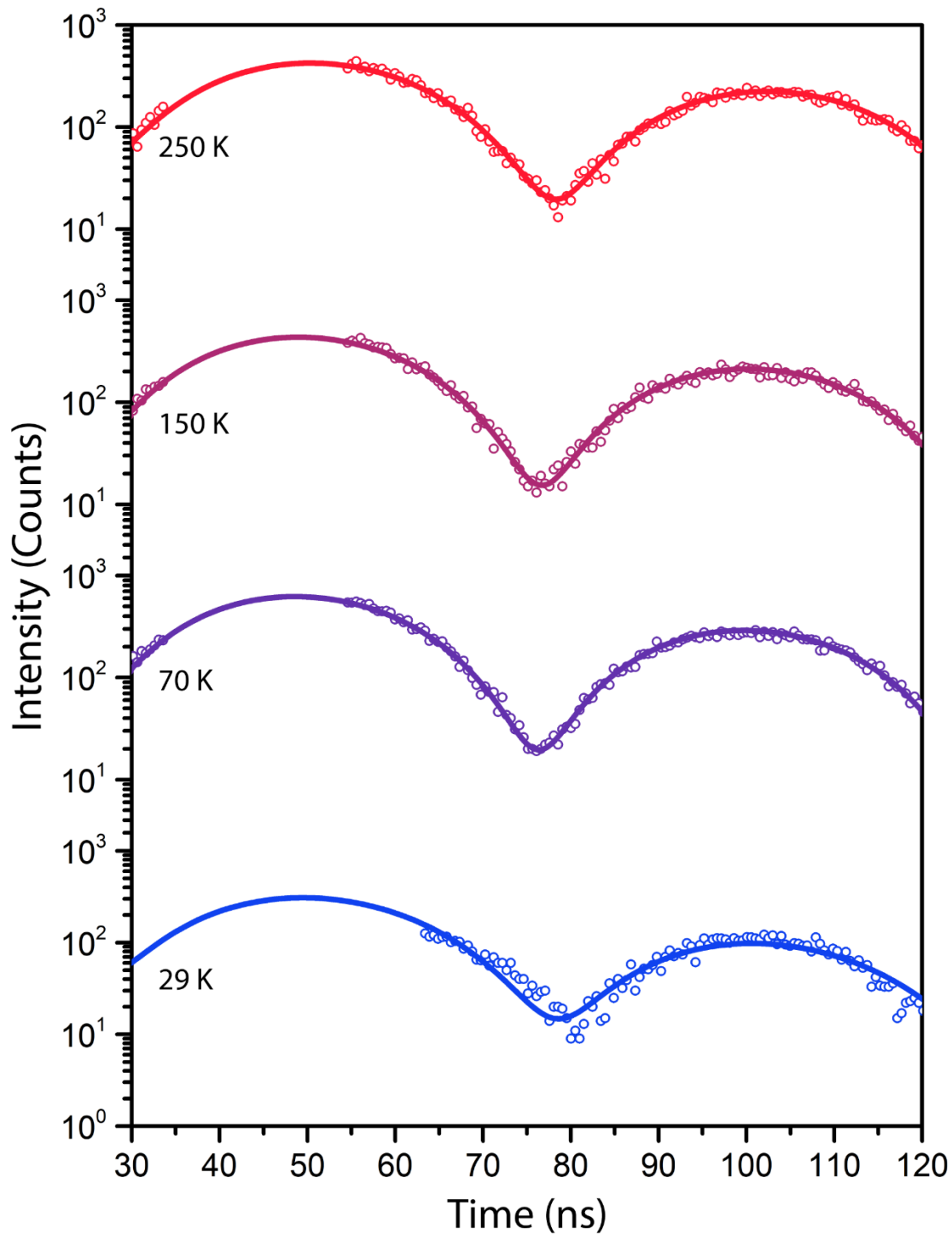

**Figure S27** | The variable temperature SMS data and fits for jarosite at ~46 GPa were collected at beamline 16-ID-D, HPCAT, APS, ANL. The temperature is specified underneath each spectrum. Spurious bunches led to non-physical features in the spectra that are masked between 34 and 54 ns, except at the lowest temperature which was masked from 20 to 63 ns. These spectra do not exhibit magnetic hyperfine splitting. The pressures and fit parameters are listed in Table S6.

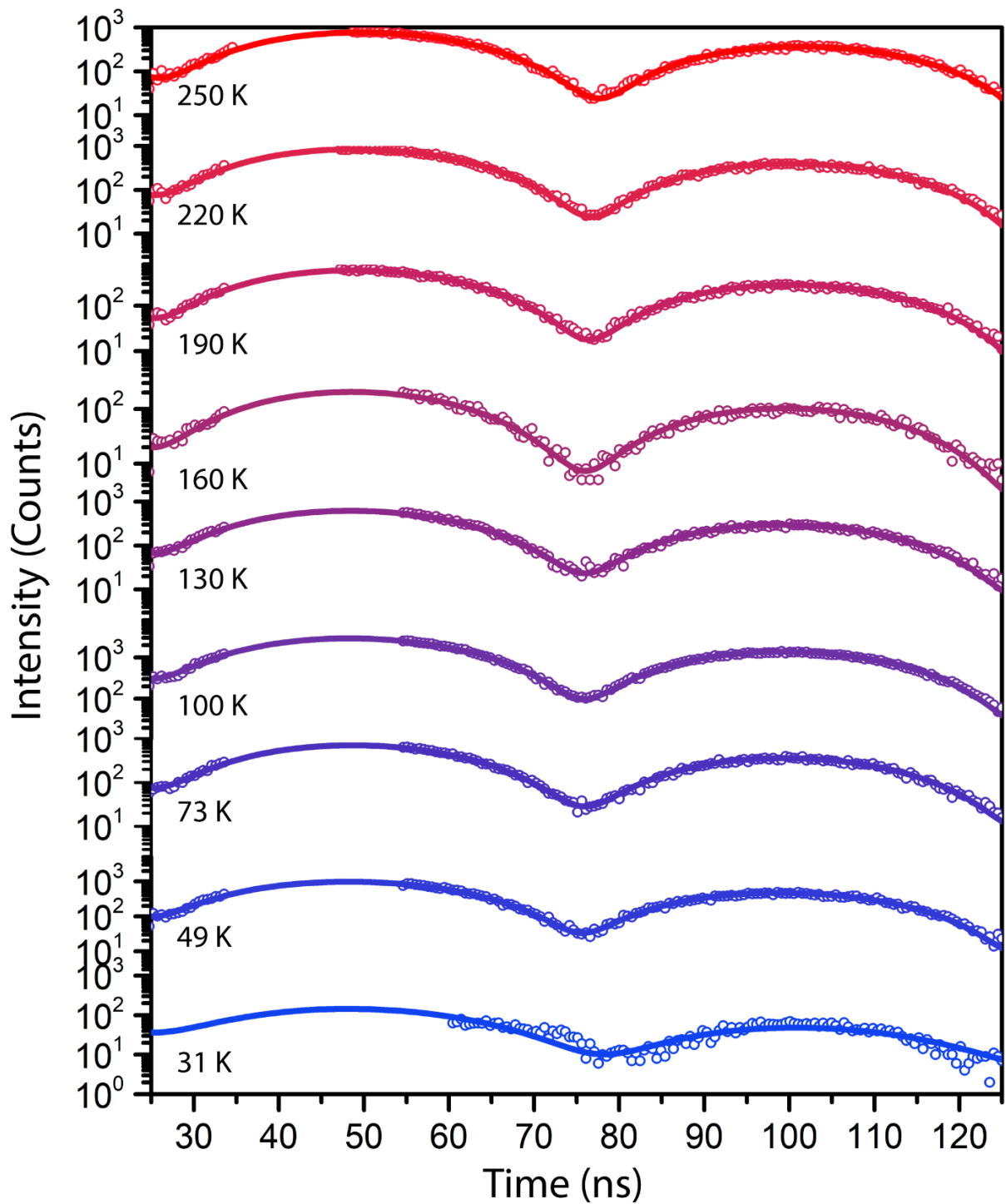

**Figure S28** | The variable temperature SMS data and fits for jarosite at ~53 GPa were collected at beamline 16-ID-D, HPCAT, APS, ANL. The temperature is specified underneath each spectrum. Spurious bunches led to non-physical features in the spectra that are masked between 34 and 54 ns, except at the lowest temperature which was masked from 20 to 60 ns. These spectra do not exhibit magnetic hyperfine splitting. The pressures and fit parameters are listed in Table S6.

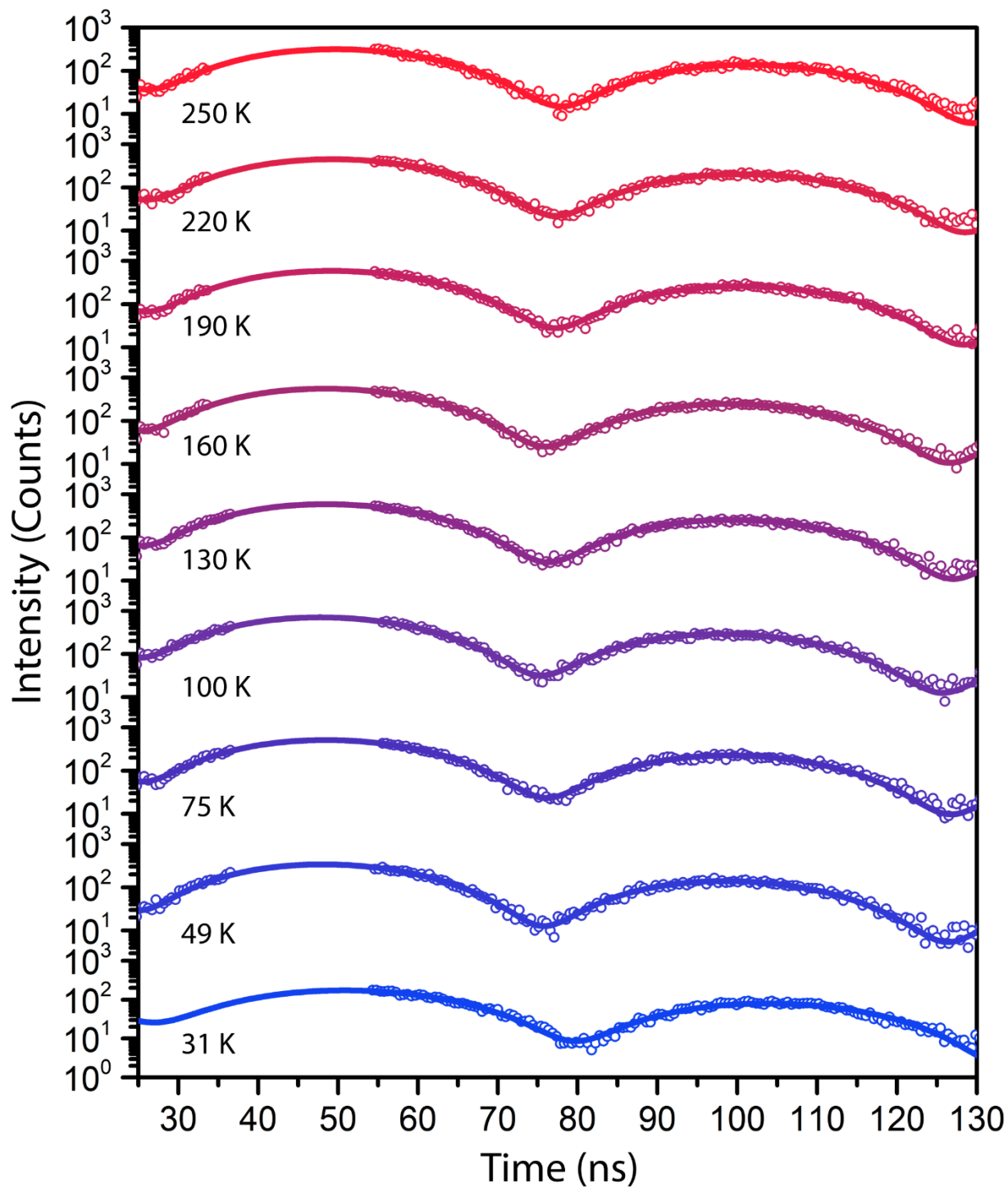

**Figure S29** | The variable temperature SMS data and fits for jarosite at ~63 GPa were collected at beamline 16-ID-D, HPCAT, APS, ANL. The temperature is specified underneath each spectrum. Spurious bunches led to non-physical features in the spectra that are masked between 37 and 55 ns, except at the lowest temperature which was masked from 20 to 54 ns. These spectra do not exhibit magnetic hyperfine splitting. The pressures and fit parameters are listed in Table S6.

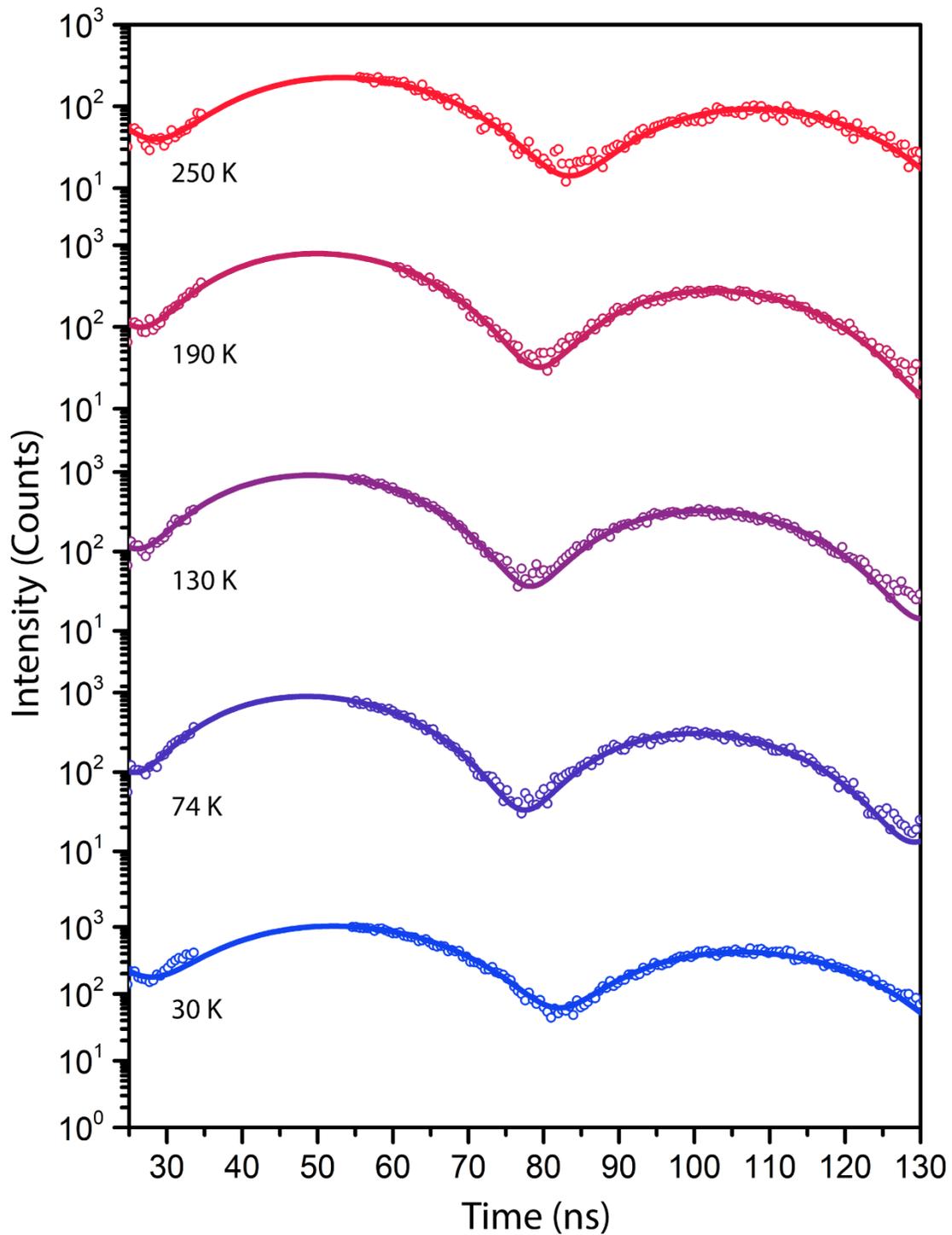

**Figure S30** | The variable temperature SMS data and fits for jarosite at ~75 GPa were collected at beamline 16-ID-D, HPCAT, APS, ANL. The temperature is specified underneath each spectrum. Spurious bunches led to non-physical features in the spectra that are masked between 34 and 54 ns. These spectra do not exhibit magnetic hyperfine splitting. The pressures and fit parameters are listed in Table S6.

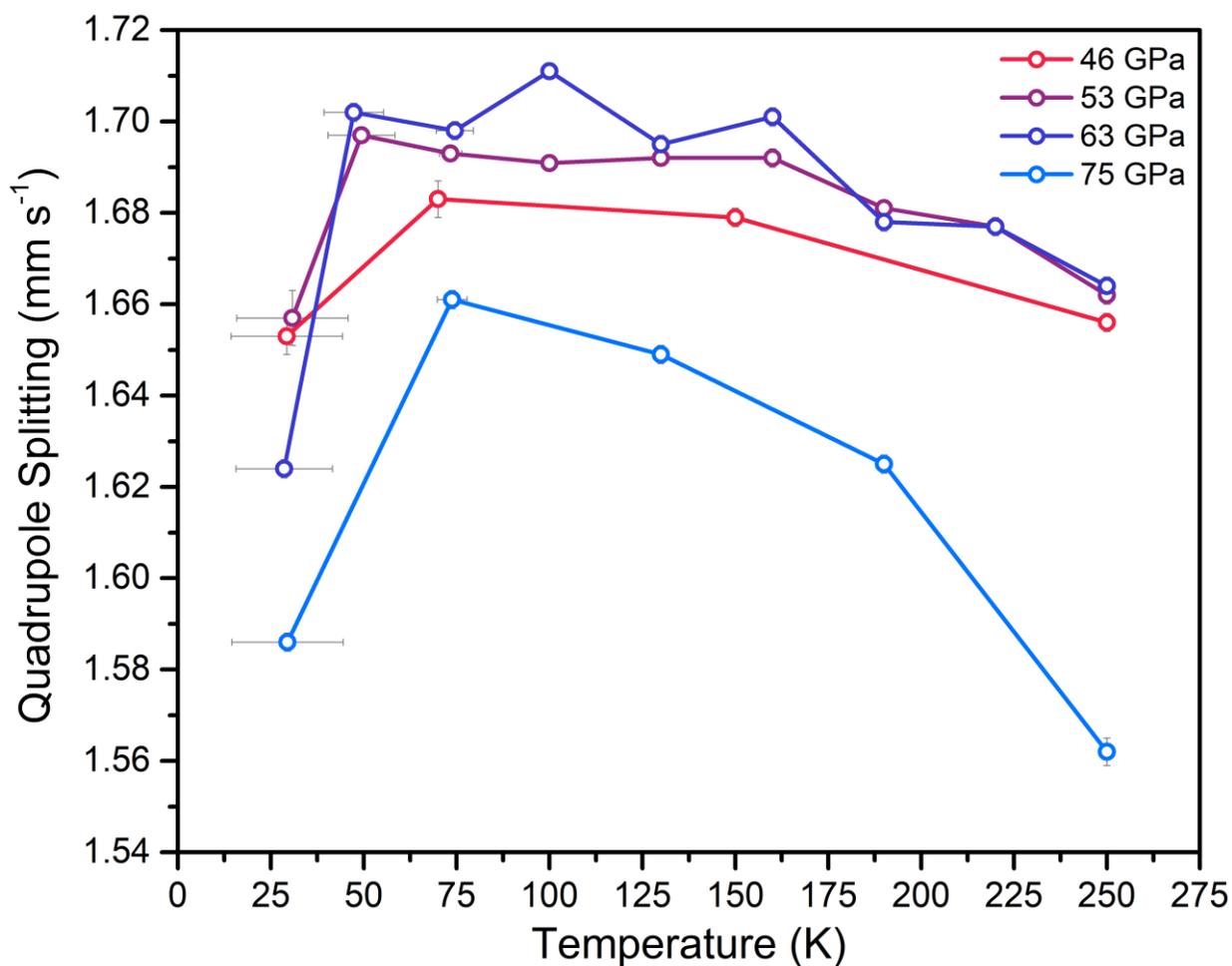

**Figure S31** | The quadrupole splitting values were extracted from the variable temperature SMS spectra at four different pressures. The splines are guides for the eye. The largest single change in $\Delta E_Q$ is about 0.07 mm s$^{-1}$. The aggregate of these data and the PXRD data collected as a function of temperature at ~65 GPa (Fig. S7) show that there is not structural or electronic phase transition in jarosite as a function of temperature between ~45 and ~75 GPa. The $\Delta E_Q$, temperature, and pressure values are given in Table S6.

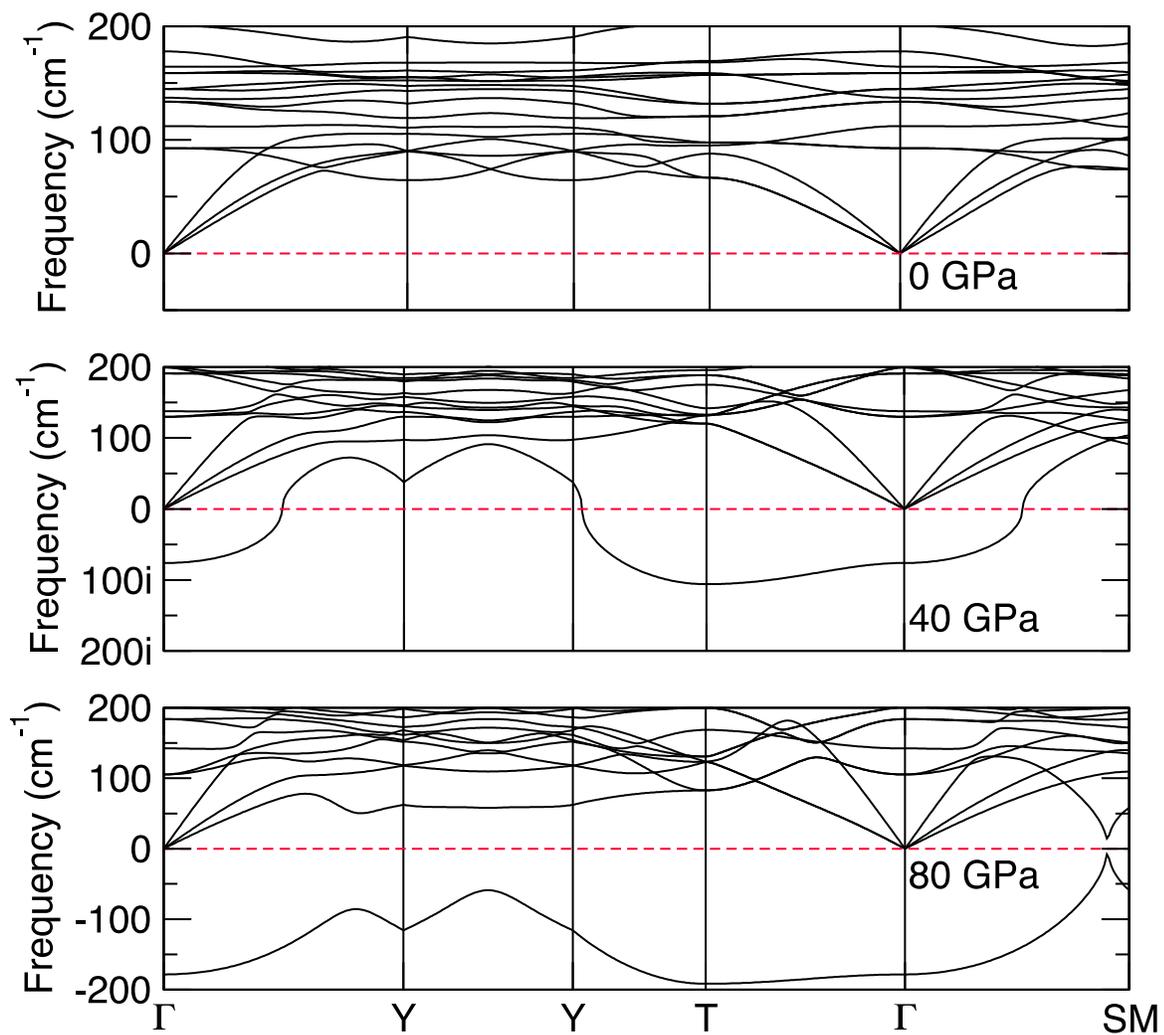

**Figure S32** | Phonon band structure at 0, 40, and 80 GPa of the $R\bar{3}m$ phase. Calculations were performed using DFT+U (5 eV), frozen-phonon technique, ferromagnetic ordering, and a 2x2x2 supercell.

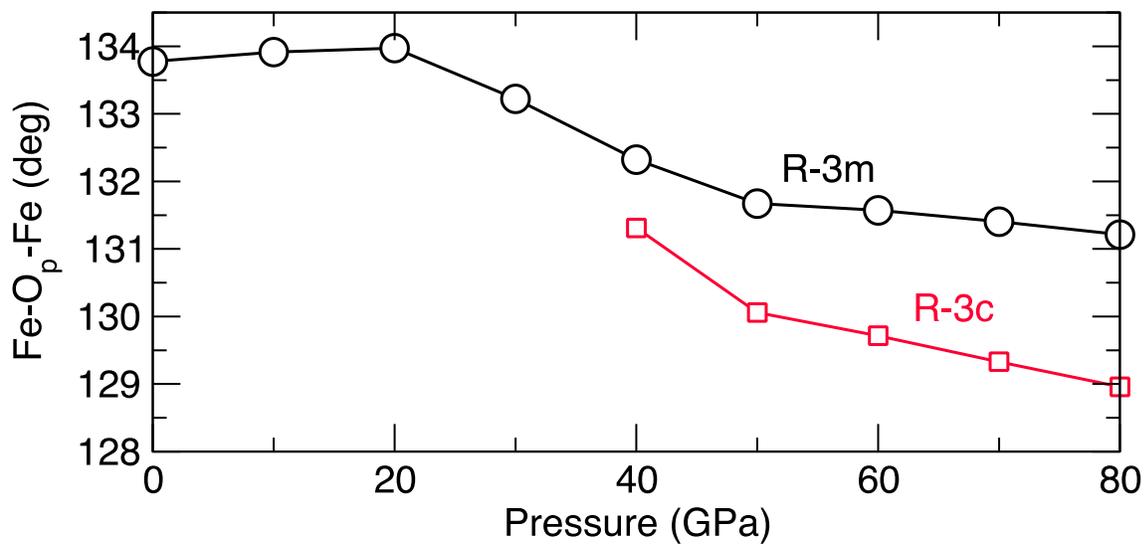

**Figure S33** | The bond angle in the Fe–O–Fe pathway within the kagomé lattice as a function of pressure based on the calculated structures.

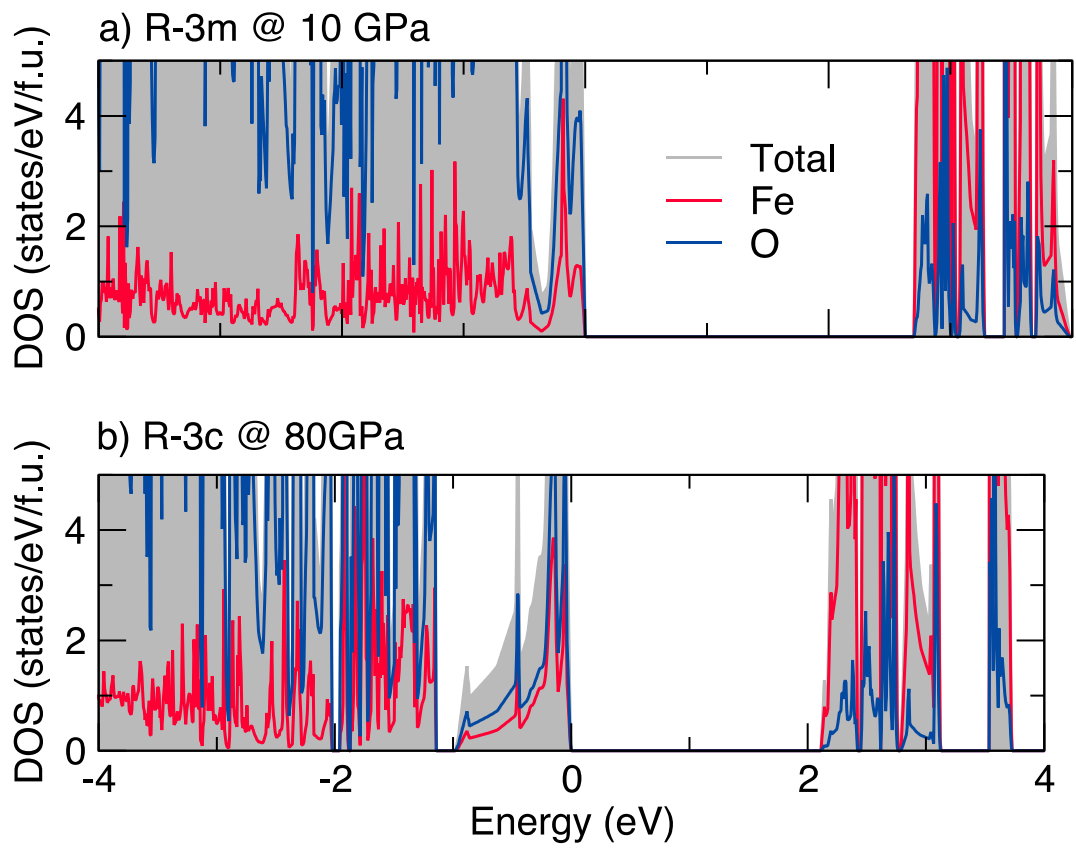

**Figure S34** | Total and projected density of states at 10 GPa (space group $R\bar{3}m$) and 80 GPa (space group $R\bar{3}c$) for the q=0 spin structure. Calculations were performed with the inclusion of spin-orbit coupling and U = 5 eV.

**Table S1** | Pressure and unit cell parameters used for the equation of state fits from 7.8(1) to 51.9(5) GPa, as obtained from ambient temperature PXRD at APS, ANL ($\lambda$ = 0.406600 Å).

| Pressure (GPa) | $a$ (Å)    | $c$ (Å)    | Volume (Å$^3$) |
|----------------|------------|------------|----------------|
| 7.8(1)         | 7.1423(3)  | 16.038(2)  | 708.5(1)       |
| 10.5(1)        | 7.1076(3)  | 15.845(2)  | 693.2(1)       |
| 12.8(1)        | 7.0846(4)  | 15.705(2)  | 682.6(1)       |
| 15.9(1)        | 7.0506(4)  | 15.532(3)  | 668.7(1)       |
| 19.5(1)        | 7.0143(5)  | 15.359(3)  | 654.4(1)       |
| 22.5(2)        | 6.9768(4)  | 15.288(3)  | 644.5(1)       |
| 25.0(2)        | 6.9461(4)  | 15.235(3)  | 636.6(1)       |
| 27.5(2)        | 6.9161(5)  | 15.178(3)  | 628.8(1)       |
| 30.1(3)        | 6.8872(5)  | 15.123(3)  | 621.2(1)       |
| 32.6(3)        | 6.8578(6)  | 15.076(3)  | 614.0(2)       |
| 33.9(3)        | 6.8418(6)  | 15.052(4)  | 610.2(2)       |
| 35.0(3)        | 6.8273(8)  | 15.029(4)  | 606.7(2)       |
| 36.0(3)        | 6.8183(7)  | 15.011(4)  | 604.4(2)       |
| 37.1(3)        | 6.8062(8)  | 14.993(5)  | 601.5(3)       |
| 38.2(3)        | 6.7918(8)  | 14.970(5)  | 598.4(3)       |
| 39.4(3)        | 6.7772(9)  | 14.964(5)  | 595.2(3)       |
| 40.6(4)        | 6.760(1)   | 14.947(6)  | 591.6(3)       |
| 41.6(4)        | 6.739(1)   | 14.940(7)  | 587.6(3)       |
| 42.7(4)        | 6.707(2)   | 14.92(1)   | 581.4(4)       |
| 43.7(4)        | 6.6752(9)  | 14.897(7)  | 574.9(3)       |
| 45.3(4)        | 6.6413(4)  | 14.764(5)  | 563.9(2)       |
| 46.0(4)        | 6.6355(3)  | 14.709(4)  | 560.0(2)       |
| 46.8(4)        | 6.6287(3)  | 14.660(4)  | 557.9(2)       |
| 47.9(4)        | 6.6215(3)  | 14.620(4)  | 555.2(1)       |
| 48.8(4)        | 6.6136(3)  | 14.582(4)  | 552.4(2)       |
| 49.9(4)        | 6.6060(3)  | 14.550(4)  | 549.9(1)       |
| 50.8(5)        | 6.5995(3)  | 14.523(3)  | 547.8(1)       |
| 51.9(5)        | 6.5932(3)  | 14.496(3)  | 545.7(1)       |

**Table S2** | Pressure and unit cell parameters used for the equation of state fits from 52.5(5) to 78.6(7) GPa, as obtained from ambient temperature PXRD at APS, ANL ($\lambda$ = 0.406600 Å).

| Pressure (GPa) | a (Å) | c (Å) | Volume (Å$^3$) |
|:---:|:---:|:---:|:---:|
| 52.5(5) | 6.5879(3) | 14.488(3) | 545.7(1) |
| 53.4(5) | 6.5829(3) | 14.471(3) | 544.6(1) |
| 54.4(5) | 6.5775(3) | 14.451(3) | 543.1(1) |
| 55.3(5) | 6.5721(3) | 14.429(3) | 541.4(1) |
| 56.3(5) | 6.5667(3) | 14.412(3) | 539.7(1) |
| 57.1(5) | 6.5618(3) | 14.393(3) | 538.2(1) |
| 58.0(5) | 6.5565(3) | 14.377(3) | 536.7(1) |
| 58.7(5) | 6.5524(3) | 14.361(3) | 535.2(1) |
| 59.5(5) | 6.5469(3) | 14.350(2) | 534.0(1) |
| 61.4(6) | 6.5402(3) | 14.331(3) | 532.6(1) |
| 62.1(6) | 6.5353(3) | 14.313(3) | 530.9(1) |
| 63.2(6) | 6.5292(3) | 14.297(3) | 529.4(1) |
| 64.2(6) | 6.5228(3) | 14.278(3) | 527.8(1) |
| 65.2(6) | 6.5177(3) | 14.267(3) | 526.1(1) |
| 66.2(6) | 6.5123(3) | 14.253(3) | 524.9(1) |
| 67.1(6) | 6.5071(3) | 14.237(3) | 523.5(1) |
| 68.1(6) | 6.5024(3) | 14.226(3) | 522.1(1) |
| 69.1(6) | 6.4975(3) | 14.214(3) | 520.9(1) |
| 69.9(6) | 6.4932(3) | 14.204(3) | 519.7(1) |
| 70.9(7) | 6.4881(3) | 14.190(3) | 518.6(1) |
| 71.7(7) | 6.4840(3) | 14.181(3) | 517.3(1) |
| 72.5(7) | 6.4795(3) | 14.174(3) | 516.3(1) |
| 73.3(7) | 6.4753(3) | 14.162(3) | 515.4(1) |
| 74.1(7) | 6.4712(3) | 14.154(3) | 514.3(1) |
| 75.5(7) | 6.4658(3) | 14.143(3) | 513.3(1) |
| 76.5(7) | 6.4605(3) | 14.131(3) | 512.1(1) |
| 77.7(7) | 6.4543(3) | 14.118(3) | 510.8(1) |
| 78.6(7) | 6.4493(4) | 14.108(3) | 509.3(1) |

**Table S3** | Pressure, temperature, and unit cell parameters for jarosite at ~65 GPa, as obtained from variable temperature PXRD measurements at APS, ANL (λ = 0.413300 Å).

| Pressure (GPa) | Temperature (K) | $a$ (Å) | $c$ (Å) | Volume (Å$^3$) |
| --- | --- | --- | --- | --- |
| 63.6(6) | 294(2) | 6.487(2) | 14.283(9) | 520.5(4) |
| 62.3(6) | 274(1) | 6.494(1) | 14.300(9) | 522.2(4) |
| 63.9(6) | 254.0(9) | 6.490(2) | 14.270(9) | 520.5(4) |
| 63.1(6) | 234.0(9) | 6.491(2) | 14.285(9) | 521.3(4) |
| 66.0(6) | 214.3(5) | 6.491(2) | 14.21(1) | 518.3(5) |
| 65.3(6) | 194.2(7) | 6.478(1) | 14.253(9) | 517.9(4) |
| 65.9(6) | 184.5(6) | 6.480(1) | 14.222(9) | 517.2(4) |
| 65.6(6) | 174.6(4) | 6.468(1) | 14.230(9) | 515.5(4) |
| 65.7(6) | 164.7(2) | 6.453(1) | 14.26(1) | 514.1(4) |
| 65.7(6) | 155.0(1) | 6.479(2) | 14.255(9) | 518.3(4) |
| 66.0(6) | 145.0(1) | 6.477(2) | 14.243(9) | 517.4(4) |
| 66.0(6) | 134.7(1) | 6.465(1) | 14.234(9) | 515.3(4) |
| 66.0(6) | 125.1(2) | 6.479(2) | 14.245(9) | 517.9(4) |
| 66.1(6) | 120.0(3) | 6.477(2) | 14.242(9) | 517.4(4) |
| 66.1(6) | 115.01(2) | 6.477(2) | 14.240(9) | 517.4(4) |
| 65.8(6) | 110.2(2) | 6.481(2) | 14.250(8) | 518.4(4) |
| 66.5(6) | 105.4(3) | 6.472(2) | 14.231(9) | 516.3(4) |
| 66.4(6) | 100.1(4) | 6.474(2) | 14.240(9) | 516.9(4) |
| 66.1(6) | 95.0(1) | 6.469(1) | 14.226(8) | 515.5(4) |
| 65.5(6) | 90.1(5) | 6.458(1) | 14.246(9) | 514.5(4) |
| 66.1(6) | 85.4(6) | 6.463(2) | 14.23(1) | 514.8(4) |
| 66.1(6) | 80.3(6) | 6.465(2) | 14.23(1) | 515.2(4) |
| 66.1(6) | 75.5(4) | 6.463(2) | 14.23(1) | 514.8(4) |
| 65.8(6) | 70.2(6) | 6.461(2) | 14.23(1) | 514.6(4) |
| 66.1(6) | 65.4(5) | 6.462(2) | 14.23(1) | 514.7(4) |
| 66.1(6) | 60.6(6) | 6.462(2) | 14.24(1) | 514.7(4) |
| 65.0(6) | 56(1) | 6.484(2) | 14.283(7) | 520.0(4) |
| 66.2(6) | 50.8(7) | 6.475(2) | 14.256(9) | 517.6(4) |
| 66.2(6) | 45.9(8) | 6.475(2) | 14.255(9) | 517.6(4) |
| 66.9(6) | 40.0(9) | 6.463(1) | 14.234(9) | 515.0(4) |
| 66.4(6) | 36(1) | 6.460(1) | 14.23(1) | 514.3(4) |
| 66.0(6) | 31(1) | 6.461(2) | 14.24(1) | 514.7(5) |
| 66.5(6) | 27(2) | 6.460(2) | 14.24(1) | 514.8(5) |
| 67.4(6) | 16(3) | 6.460(2) | 14.23(1) | 514.5(4) |

**Table S4** | Third-order Birch–Murnaghan equation of state parameters for unit cell volume. The low-pressure range fits 7.8(1) to 15.9(1) GPa. The mid-pressure range fits 19.5(1) to 39.4(3) GPa. The high-pressure range fits 49.9(5) to 78.6(7) GPa.

| Unit cell volume | | | | |
|---|---|---|---|---|
| | Low Pressure | Mid Pressure | High Pressure | Entire Pressure Range |
| $V_0$ (Å$^3$) | 771(7) | 729(2) | 762(2) | |
| $K_0$ (GPa) | 68(14) | 163.2(5) | 52.3(6) | |
| $K_0'$ | 7(2) | 2.09(3) | 6.74(9) | |
| weighted-$\chi^2$ | 4.78 | 1.02 | 2.07 | 729.92 |

| $a$ unit cell parameter | | | | |
|---|---|---|---|---|
| | Low Pressure | Mid Pressure | High Pressure | Entire Pressure Range |
| $L_0$ (Å$^3$) | 7.27(3) | 7.276(5) | 7.05(3) | |
| $M_0$ (GPa) | 353(168) | 491(13) | 524(60) | |
| $M_0'$ | 27(22) | 5.2(3) | 12(1) | |
| weighted-$\chi^2$ | 16.16 | 1.37 | 1.55 | 1491.0 |

| $c$ unit cell parameter | | | | |
|---|---|---|---|---|
| | Low Pressure | Mid Pressure | High Pressure | Entire Pressure Range |
| $L_0$ (Å$^3$) | 16.86(6) | 16.00(7) | 17.68(2) | |
| $M_0$ (GPa) | 111(15) | 356(17) | 47(3) | |
| $M_0'$ | 14(2) | 15(3) | 23.3 | |
| weighted-$\chi^2$ | 0.23 | 0.97 | 5.53 | 224.33 |

**Table S5** | Synchrotron Mössbauer spectroscopy fit parameters at ambient temperature. The SMS data point obtained at the transition pressure has two centers because of the transition. The two data points obtained as the highest pressure in each of the respective experiments (i.e. at 72.3(7) and at 121(7) GPa) have two centers because of the strain. Modeling these data with a distribution in one parameter alone is insufficient because the pressure gradient across the sample and the strain at the sample create a gradient in all relevant fitting variables, not just in $\Delta E_Q$ alone. Data modeled with two sites are denoted by a * next to the measurement pressure. The relative weights of these sites are given.

| Pressure (GPa) | $\Delta E_Q$ (mm s$^{-1}$) | $B_{HF}$ (T) | $\Delta E_Q$ distribution (mm s$^{-1}$) | Weight (%) | $\chi^2$ |
|---|---|---|---|---|---|
| 35.9(3) | 2.19(2) | 0 | 0.15 | | 1.80 |
| 39.5(4) | 2.13(2) | 0 | 0.18 | | 1.88 |
| 45.6(4) | 2.24(2) | 0 | 0.92 | 54 | 1.87 |
| 45.6(4) * | 1.62(2) | 0 | 0.16 | 46 | 1.87 |
| 51.3(5) | 1.65(2) | 0 | 0.02 | | 2.34 |
| 57.6(5) | 1.66(2) | 0 | 0.08 | | 1.64 |
| 64.1(6) | 1.67(2) | 0 | 0.14 | | 1.44 |
| 72.3(7) | 1.58(2) | 0 | | 25 | 2.30 |
| 72.3(7)* | 1.744(1) | 0 | | 75 | 2.30 |
| 88(4) | 1.57(1) | 0 | | | 1.14 |
| 99(5) | 1.47(1) | 0 | | | 1.43 |
| 111(5) | 1.39(1) | 0 | | | 3.19 |
| 121(7) | 1.23(1) | 0 | | 96 | 1.37 |
| 121(7) * | 1.93(2) | 0 | | 4.0 | 1.37 |

**Table S6** | Synchrotron Mössbauer fit parameters at variable temperature and pressure conditions. Error in temperature is estimated based on the discrepancy between the measured temperatures from the two thermocouples. Therefore, the error is much larger than the precision given for the low temperature measurements.

| Pressure (GPa) | Temperature (K) | $\Delta E_Q$ (mm s$^{-1}$) | $B_{HF}$ (T) | $\chi^2$ |
|---|---|---|---|---|
| 45.5(5) | 250(1) | 1.656(2) | 0 | 1.82 |
| 46.7(5) | 150(1) | 1.679(2) | 0 | 1.67 |
| 45.9(5) | 70(2) | 1.683(4) | 0 | 1.56 |
| 46.7(5) | 29.3(15) | 1.653(4) | 0 | 2.90 |
| 49.6(5) | 250(1) | 1.662(2) | 0 | 2.06 |
| 51.1(5) | 220(1) | 1.677(1) | 0 | 1.76 |
| 52.6(5) | 190(1) | 1.681(1) | 0 | 1.51 |
| 53.0(5) | 160(1) | 1.692(2) | 0 | 1.82 |
| 53.4(5) | 130(1) | 1.692(1) | 0 | 1.50 |
| 53.4(5) | 100(1) | 1.6909(9) | 0 | 2.84 |
| 52.2(5) | 73.4(3) | 1.693(1) | 0 | 1.51 |
| 52.2(5) | 49.4(9) | 1.697(1) | 0 | 2.47 |
| 52.6(5) | 30.8(15) | 1.657(6) | 0 | 2.51 |
| 63.3(6) | 250(1) | 1.664(2) | 0 | 1.47 |
| 63.7(6) | 220(1) | 1.677(2) | 0 | 1.80 |
| 63.0(6) | 190(1) | 1.678(2) | 0 | 1.80 |
| 62.5(6) | 160(1) | 1.701(2) | 0 | 1.73 |
| 63.7(6) | 130(1) | 1.695(2) | 0 | 1.99 |
| 62.2(6) | 100(1) | 1.711(1) | 0 | 1.89 |
| 63.7(6) | 74.5(5) | 1.698(2) | 0 | 1.69 |
| 63.3(6) | 47.3(8) | 1.702(2) | 0 | 1.07 |
| 63.7(6) | 28.6(13) | 1.624(2) | 0 | 1.20 |
| 78.3(7) | 250(1) | 1.562(3) | 0 | 1.14 |
| 75.2(7) | 190(1) | 1.625(2) | 0 | 1.58 |
| 74.4(7) | 130(1) | 1.649(2) | 0 | 1.86 |
| 75.6(7) | 73.8(4) | 1.661(2) | 0 | 1.60 |
| 74.0(7) | 29.5(15) | 1.586(2) | 0 | 3.55 |

**Table S7** | Summary of Rietveld refinement results using the $R\bar{3}m$, $R\bar{3}$, and $R\bar{3}c$ models for the 62.1(6) GPa PXRD pattern. Values in parenthesis represent one standard deviation.

| | | | | | | |
|---|---|---|---|---|---|---|
| $R\bar{3}c$ | $R_{exp}$ | 2.208 | $R_{wp}$ | 0.886 | $R_p$ | 0.636 |
| | $a$ (Å) | 6.5459(7) | $c$ (Å) | 28.77(1) | $V$ (Å³) | 1067.8(5) |
| | **Atom** | **x** | **y** | **z** | **Occ** | **Beq** |
| | Fe | 0.52788 | 0.00000 | 0.25000 | 1 | 0.5 |
| | K | 0.00000 | 0.00000 | 0.00000 | 1 | 0.5 |
| | S | 0.00000 | 0.00000 | 0.16095 | 1 | 0.5 |
| | O1 | 0.00000 | 0.00000 | 0.21203 | 1 | 0.5 |
| | O2 | 0.77199 | 0.80781 | 0.14731 | 1 | 0.5 |
| | O3 | 0.59339 | 0.77557 | 0.23051 | 1 | 0.5 |
| | H | 0.76432 | 0.8526 | 0.22382 | 1 | 0.5 |
| | Spherical harmonics preferred orientation terms | | | | | |
| | $y_{00}$ | 1 | $y_{40}$ | 0.88(4) | | |
| | $y_{20}$ | −1.20(2) | $y_{43m}$ | 0.03(1) | | |
| $R\bar{3}m$ | $R_{exp}$ | 2.207 | $R_{wp}$ | 1.464 | $R_p$ | 0.880 |
| | $a$ (Å) | 6.539(1) | $c$ (Å) | 14.42(1) | $V$ (Å³) | 533.8(6) |
| | **Atom** | **x** | **y** | **z** | **Occ** | **Beq** |
| | Fe | 0.16667 | −0.16667 | −0.16667 | 1 | 0.5 |
| | K | 0.00000 | 0.00000 | 0.00000 | 1 | 0.5 |
| | S | 0.00000 | 0.00000 | 0.30770 | 1 | 0.5 |
| | O1 | 0.00000 | 0.00000 | 0.39130 | 1 | 0.5 |
| | O2 | 0.22320 | −0.22320 | −0.05488 | 1 | 0.5 |
| | O3 | 0.12731 | −0.12731 | 0.13499 | 1 | 0.5 |
| | H | 0.19585 | −0.19585 | 0.10988 | 1 | 0.5 |
| | Spherical harmonics preferred orientation terms | | | | | |
| | $y_{00}$ | 1 | $y_{40}$ | 0.88(4) | | |
| | $y_{20}$ | −1.66(4) | $y_{43m}$ | −0.22(2) | | |
| $R\bar{3}$ | $R_{exp}$ | 2.204 | $R_{wp}$ | 0.689 | $R_p$ | 0.507 |
| | $a$ (Å) | 6.5459(5) | $c$ (Å) | 14.374(4) | $V$ (Å³) | 533.4(2) |
| | **Atom** | **x** | **y** | **z** | **Occ** | **Beq** |
| | Fe | 0.50000 | 0.00000 | 0.50000 | 1 | 0.5 |
| | K | 0.00000 | 0.00000 | 0.00000 | 1 | 0.5 |
| | S | 0.00000 | 0.00000 | 0.32264 | 1 | 0.5 |
| | O1 | 0.00000 | 0.00000 | 0.42483 | 1 | 0.5 |
| | O2 | 0.81952 | 0.76647 | 0.29590 | 1 | 0.5 |
| | O3 | 0.12872 | 0.26307 | 0.13049 | 1 | 0.5 |
| | Spherical harmonics preferred orientation terms | | | | | |
| | $y_{00}$ | 1 | $y_{40}$ | 0.89(2) | | |
| | $y_{20}$ | −1.32(2) | $y_{43p}$ | −1.11(1) | | |
| | | | $y_{43m}$ | −0.87(6) | | |